\documentclass[runningheads]{llncs}

\usepackage{booktabs}   \usepackage{caption} 
\usepackage{subcaption}
\usepackage{bussproofs}
\usepackage[cal=boondoxo]{mathalfa}
\DeclareMathAlphabet{\mathpzc}{OT1}{pzc}{m}{it}
\usepackage{amsmath}
\usepackage{amssymb}
\usepackage{color}
\usepackage{xspace}

\newcommand{\ignore}[1]{{}}

\newcommand{\syntaxDef}[3]{\rulebox{\syntaxKeyword$#1\mathrel{::=}{#2}$ \ifthenelse{\equal{#3}{}}{}{[#3]}}}

\makeatletter
\newcommand{\shorteq}{\settowidth{\@tempdima}{-}\resizebox{\@tempdima}{\height}{=}}
\makeatother

  \newcommand{\dom}[1]{\ensuremath{\mathtt{dom}(#1)}}

 \usepackage{appendix}

\usepackage{xcolor}
\usepackage{colortbl}

\usepackage[utf8]{inputenc}

\usepackage{stmaryrd}
\usepackage{hyperref}

\usepackage{wrapfig}

\usepackage{float}
\usepackage{balance}

\usepackage{textcomp} 

\usepackage{changepage}
\usepackage{array,etoolbox}

\preto\tabular{\setcounter{magicrownumbers}{0}}
\newcounter{magicrownumbers}

\usepackage{cleveref} 

\usepackage{wrapfig}
\usepackage{caption}

\usepackage{longtable}
\usepackage{tabu}

\usepackage{wasysym}
\usepackage{mathpartir}

\usepackage{mathtools} \usepackage{xfrac}

\usepackage[qm,braket]{qcircuit}

\newcommand{\disq}{\textsc{DisQ}\xspace}

\newcommand{\qass}[2]{{#1}\;{\leftarrow}\;{#2}}

\newcommand{\cglocus}[2]{\textcolor{spec}{\langle #1\rangle_{#2}}}
\newcommand{\glocus}[2]{\langle #1\rangle_{#2}}
\newcommand{\bglocus}[2]{\big{\langle} #1\big{\rangle}_{#2}}
\newcommand{\sif}[3]{\texttt{if}~\cn{(}{#1}\cn{)}~{#2}~{\cn{else}}~{#3}}

\newcommand{\mmod}{\cn{\%}}

\colorlet{kwd}{black!80!green}
\definecolor{spec1}{RGB}{78, 131, 162}
\definecolor{spec0}{RGB}{66, 102, 136}
\definecolor{lespec}{RGB}{30, 80, 180}
\colorlet{spec}{lespec}
\colorlet{auto}{lespec!35!lightgray}
\colorlet{stack}{magenta}

\newcommand{\qafny}{\rulelab{Qafny}\xspace}

\newcommand{\myparagraph}[1]{\noindent\paragraph{\textbf{#1}}}

\usepackage{tikz}

\newcommand{\cmode}{\texttt{C}}

\newcommand{\qmodename}{\texttt{Q}}
\newcommand{\qmode}[1]{\texttt{Q}(#1)}

\newcommand{\slen}[1]{|#1|}

\usepackage{bbold}
\newcommand{\zero}{\mathbb{0}}

\newcommand{\CC}{\ensuremath{\mathbb{C}}\xspace}

\usepackage{pgf}

\usepackage{tikz} \usetikzlibrary{arrows,shapes.misc,shapes.geometric, shapes.arrows,shapes.callouts,
  shapes.gates.logic.US,
  chains,matrix,positioning,scopes,decorations.pathmorphing,decorations.text,
  decorations.pathreplacing, shadows,automata,
  fit, calc, arrows.meta
}

\tikzset{ machine/.style={
rectangle,
minimum width=25mm,
    minimum height=18mm,
    text width=24mm,
align=center,
very thick,
    draw=black,
color=black,
    fill=white,
}
}

\newcommand\wideparen[1]{\tikz[baseline=(wideArcAnchor.base)]{
    \node[inner sep=0] (wideArcAnchor) {$#1$}; 
    \coordinate (wideArcAnchorA) at ($0.9*(wideArcAnchor.north west) + 0.1*(wideArcAnchor.north east)+(0.0em,0.75ex)$);
    \coordinate (wideArcAnchorB) at ($0.1*(wideArcAnchor.north west) + 0.9*(wideArcAnchor.north east)+(0.0em,0.75ex)$);
\draw[line width=0.1ex,line cap=round] 
        ($(wideArcAnchor.north west)+(0.0em,0.1ex)$) 
            .. controls (wideArcAnchorA) and (wideArcAnchorB) ..
        ($(wideArcAnchor.north east)+(0.0em,0.1ex)$)        
    ;
}}
\newcommand{\cmsg}[1]{\wideparen{#1}}

\DeclarePairedDelimiter\abs{\lvert}{\rvert}
\DeclarePairedDelimiter\norm{\lVert}{\rVert}

\makeatletter
\let\oldabs\abs
\def\abs{\@ifstar{\oldabs}{\oldabs*}}
\let\oldnorm\norm
\def\norm{\@ifstar{\oldnorm}{\oldnorm*}}
\makeatother

\makeatletter
\DeclareRobustCommand{\vardivision}{\mathbin{\mathpalette\@vardivision\relax}}
\newcommand{\@vardivision}[2]{\reflectbox{$\m@th\smallsetminus$}}
\makeatother

\DeclareFontFamily{U} {MnSymbolC}{}
\DeclareFontShape{U}{MnSymbolC}{m}{n}{
  <-6> MnSymbolC5
  <6-7> MnSymbolC6
  <7-8> MnSymbolC7
  <8-9> MnSymbolC8
  <9-10> MnSymbolC9
  <10-12> MnSymbolC10
  <12-> MnSymbolC12}{}
\DeclareFontShape{U}{MnSymbolC}{b}{n}{
  <-6> MnSymbolC-Bold5
  <6-7> MnSymbolC-Bold6
  <7-8> MnSymbolC-Bold7
  <8-9> MnSymbolC-Bold8
  <9-10> MnSymbolC-Bold9
  <10-12> MnSymbolC-Bold10
  <12-> MnSymbolC-Bold12}{}

\DeclareSymbolFont{MnSyC} {U} {MnSymbolC}{m}{n}

\DeclareMathSymbol{\sqcupplus}{\mathbin}{MnSyC}{70}

\usepackage{booktabs}

\usepackage{alltt}
\usepackage{listings,lstcoq}
\definecolor{ltblue}{rgb}{0,0.4,0.4}
\definecolor{dkblue}{rgb}{0,0.1,0.6}
\definecolor{dkgreen}{rgb}{0,0.35,0}
\definecolor{dkviolet}{rgb}{0.3,0,0.5}
\definecolor{dkred}{rgb}{0.5,0,0}
\usepackage[export]{adjustbox}

\newcommand{\code}[1]{{\small\texttt{#1}}}

\newcommand{\rulelab}[1]{{\small \textsc{#1}}}

\newcommand{\sact}[2]{#1\,\textbf{.}\,#2}

\newcommand{\sifb}[3]{\texttt{if}~{(#1)}~{#2}~\texttt{else}~{#3}}
\newcommand{\sifc}[2]{\texttt{if}~{(#1)}~\{{#2}\}}

\newcommand{\csenda}[2]{#1\texttt{!}#2}

\newcommand{\creva}[2]{#1\texttt{?(}#2\texttt{)}}

\newcommand{\seq}[2]{#1 \cn{.} #2}
\newcommand{\sdot}[2]{#1\texttt{,}\,#2}
\newcommand{\sacell}[2]{#1\textcolor{teal}{\{\!|} #2 \textcolor{teal}{|\!\}}}
\newcommand{\scell}[1]{\textcolor{teal}{\{\!|} #1 \textcolor{teal}{|\!\}}}
\newcommand{\bsacell}[2]{#1\textcolor{teal}{\big{\{}\!\big{|}} #2 \textcolor{teal}{\big{|}\!\big{\}}}}
\newcommand{\bscell}[1]{\textcolor{teal}{\big{\{}\!\big{|}} #1 \textcolor{teal}{\big{|}\!\big{\}}}}

\newcommand{\downc}[2]{\partial \,#1 (#2)}

\newcommand{\downd}[1]{\partial \,#1}

\newcommand{\tnort}{\texttt{Nor}}

\newcommand{\thadt}{\texttt{Had}}

\newcommand{\tcht}{\texttt{EN}}
\newcommand{\shad}[3]{\frac{1}{\sqrt{#1}}\Motimes_{j=0}^{#2}{(\ket{0}+#3\ket{1})}}

\newcommand{\ssum}[3]{\Msum_{#1}^{#2}{#3}}
\newcommand{\sch}[3]{\Msum_{j=0}^{#1}{#2 {#3}}}
\newcommand{\scha}[4]{\Msum_{j=0}^{#1}{#2\ket{#3}#4}}
\newcommand{\schai}[4]{\Msum_{i=0}^{#1}{#2\ket{#3}#4}}

\newcommand{\schk}[3]{\Msum_{k=0}^{#1}{#2\ket{#3}}}
\newcommand{\schii}[3]{\Msum_{i=0}^{#1}{#2\ket{#3}}}

\newcommand{\schi}[3]{\Msum^{#1}{#2\ket{#3}}}

\DeclareMathOperator*{\Motimes}{\text{\raisebox{0.25ex}{\scalebox{0.8}{$\bigotimes$}}}}
\DeclareMathOperator*{\Msum}{\text{\raisebox{0.25ex}{\scalebox{0.8}{$\sum$}}}}

\newcommand{\cupdot}{\mathbin{\mathaccent\cdot\cup}}

\newcommand{\ssassign}[3]{{#1} \xleftarrow{#2} {#3}}

\newcommand{\smea}[2]{{#1}\leftarrow {\mathpzc{M}\cn{(}#2\cn{)}}}

\newcommand{\sminus}{\texttt{-}}
\newcommand{\splus}{\texttt{+}}

\newcommand{\srange}[2]{[#1,#2)}

\newcommand{\cn}[1]{\texttt{#1}}

\newcommand{\denote}[1]{\llbracket #1 \rrbracket\xspace}

\newcommand{\tob}[1]{[\!\!( #1 )\!\!]}
\newcommand{\qfun}[2]{#1\langle #2 \rangle}

\usepackage{newunicodechar}
\let\Alpha=A
\let\Beta=B
\let\Epsilon=E
\let\Zeta=Z
\let\Eta=H
\let\Iota=I
\let\Kappa=K
\let\Mu=M
\let\Nu=N
\let\Omicron=O
\let\omicron=o
\let\Rho=P
\let\Tau=T
\let\Chi=X

\newunicodechar{Α}{\ensuremath{\Alpha}}
\newunicodechar{α}{\ensuremath{\alpha}}
\newunicodechar{Β}{\ensuremath{\Beta}}
\newunicodechar{β}{\ensuremath{\beta}}
\newunicodechar{Γ}{\ensuremath{\Gamma}}
\newunicodechar{γ}{\ensuremath{\gamma}}
\newunicodechar{Δ}{\ensuremath{\Delta}}
\newunicodechar{δ}{\ensuremath{\delta}}
\newunicodechar{Ε}{\ensuremath{\Epsilon}}
\newunicodechar{ε}{\ensuremath{\epsilon}}
\newunicodechar{ϵ}{\ensuremath{\varepsilon}}
\newunicodechar{Ζ}{\ensuremath{\Zeta}}
\newunicodechar{ζ}{\ensuremath{\zeta}}
\newunicodechar{Η}{\ensuremath{\Eta}}
\newunicodechar{η}{\ensuremath{\eta}}
\newunicodechar{Θ}{\ensuremath{\Theta}}
\newunicodechar{θ}{\ensuremath{\theta}}
\newunicodechar{ϑ}{\ensuremath{\vartheta}}
\newunicodechar{Ι}{\ensuremath{\Iota}}
\newunicodechar{ι}{\ensuremath{\iota}}
\newunicodechar{Κ}{\ensuremath{\Kappa}}
\newunicodechar{κ}{\ensuremath{\kappa}}
\newunicodechar{Λ}{\ensuremath{\Lambda}}
\newunicodechar{λ}{\ensuremath{\lambda}}
\newunicodechar{Μ}{\ensuremath{\Mu}}
\newunicodechar{μ}{\ensuremath{\mu}}
\newunicodechar{Ν}{\ensuremath{\Nu}}
\newunicodechar{ν}{\ensuremath{\nu}}
\newunicodechar{Ξ}{\ensuremath{\Xi}}
\newunicodechar{ξ}{\ensuremath{\xi}}
\newunicodechar{Ο}{\ensuremath{\Omicron}}
\newunicodechar{ο}{\ensuremath{\omicron}}
\newunicodechar{Π}{\ensuremath{\Pi}}
\newunicodechar{π}{\ensuremath{\pi}}
\newunicodechar{ϖ}{\ensuremath{\varpi}}
\newunicodechar{Ρ}{\ensuremath{\Rho}}
\newunicodechar{ρ}{\ensuremath{\rho}}
\newunicodechar{ϱ}{\ensuremath{\varrho}}
\newunicodechar{Σ}{\ensuremath{\Sigma}}
\newunicodechar{σ}{\ensuremath{\sigma}}
\newunicodechar{ς}{\ensuremath{\varsigma}}
\newunicodechar{Τ}{\ensuremath{\Tau}}
\newunicodechar{τ}{\ensuremath{\tau}}
\newunicodechar{Υ}{\ensuremath{\Upsilon}}
\newunicodechar{υ}{\ensuremath{\upsilon}}
\newunicodechar{Φ}{\ensuremath{\Phi}}
\newunicodechar{φ}{\ensuremath{\phi}}
\newunicodechar{ϕ}{\ensuremath{\varphi}}
\newunicodechar{Χ}{\ensuremath{\Chi}}
\newunicodechar{χ}{\ensuremath{\chi}}
\newunicodechar{Ψ}{\ensuremath{\Psi}}
\newunicodechar{ψ}{\ensuremath{\psi}}
\newunicodechar{Ω}{\ensuremath{\Omega}}
\newunicodechar{ω}{\ensuremath{\omega}}

\newunicodechar{ℕ}{\ensuremath{\mathbb{N}}}
\newunicodechar{∅}{\ensuremath{\emptyset}}

\newunicodechar{∙}{\ensuremath{\bullet}}
\newunicodechar{≈}{\ensuremath{\approx}}
\newunicodechar{≅}{\ensuremath{\cong}}
\newunicodechar{≡}{\ensuremath{\equiv}}
\newunicodechar{≤}{\ensuremath{\le}}
\newunicodechar{≥}{\ensuremath{\ge}}
\newunicodechar{≠}{\ensuremath{\neq}}
\newunicodechar{∀}{\ensuremath{\forall}}
\newunicodechar{∃}{\ensuremath{\exists}}
\newunicodechar{±}{\ensuremath{\pm}}
\newunicodechar{∓}{\ensuremath{\pm}}
\newunicodechar{·}{\ensuremath{\cdot}}
\newunicodechar{⋯}{\ensuremath{\cdots}}
\newunicodechar{…}{\ensuremath{\ldots}}
\newunicodechar{∷}{~\mathrel{:\!\!\!:}~}
\newunicodechar{×}{\ensuremath{\times}}
\newunicodechar{∞}{\ensuremath{\infty}}
\newunicodechar{→}{\ensuremath{\to}}
\newunicodechar{←}{\ensuremath{\leftarrow}}
\newunicodechar{⇒}{\ensuremath{\Rightarrow}}
\newunicodechar{↦}{\ensuremath{\mapsto}}
\newunicodechar{↝}{\ensuremath{\leadsto}}
\newunicodechar{∨}{\ensuremath{\vee}}
\newunicodechar{∧}{\ensuremath{\wedge}}
\newunicodechar{⊢}{\ensuremath{\vdash}}
\newunicodechar{⊣}{\ensuremath{\dashv}}
\newunicodechar{∣}{\ensuremath{\mid}}
\newunicodechar{∈}{\ensuremath{\in}}
\newunicodechar{⊆}{\ensuremath{\subseteq}}
\newunicodechar{⊂}{\ensuremath{\subset}}
\newunicodechar{∪}{\ensuremath{\cup}}
\newunicodechar{⋓}{\ensuremath{\Cup}}
\newunicodechar{∉}{\ensuremath{\not\in}}
\newunicodechar{√}{\ensuremath{\sqrt}}

\newunicodechar{⊸}{\ensuremath{\multimap}}
\newunicodechar{⊗}{\ensuremath{\otimes}}
\newunicodechar{⨂}{\ensuremath{\bigotimes}}
\newunicodechar{⊕}{\ensuremath{\oplus}}
\newunicodechar{〈}{\ensuremath{\langle}}
\newunicodechar{⟨}{\ensuremath{\langle}}
\newunicodechar{⟩}{\ensuremath{\rangle}}
\newunicodechar{〉}{\ensuremath{\rangle}}
\newunicodechar{¡}{\ensuremath{\upsidedownbang}}
\newunicodechar{∘}{\ensuremath{\circ}}
\newunicodechar{†}{\ensuremath{\dagger}}
\newunicodechar{⊤}{\ensuremath{\top}}
\newunicodechar{⊥}{\ensuremath{\bot}}

\newunicodechar{〚}{\ensuremath{\llbracket}}
\newunicodechar{〛}{\ensuremath{\rrbracket}}

\usepackage{etoolbox} \newtoggle{comments}
\toggletrue{comments}
\usepackage[normalem]{ulem}

\iftoggle{comments}{
  \newcommand{\fixme}[1]{\textbf{\textcolor{red}{[ Fixme: #1]}}}
  \newcommand{\todo}[1]{\textbf{\textcolor{green}{[ TODO: #1 ]}}}
  \newcommand{\mwh}[1]{\textbf{\textcolor{red}{[ Mike: #1 ]}}}
  
  \newcommand{\khh}[1]{\textbf{\textcolor{orange}{[ Kesha: #1 ]}}}
  \newcommand{\shh}[1]{\textbf{\textcolor{purple}{[ Shih-Han: #1 ]}}}
  \newcommand{\liyi}[1]{\textbf{\textcolor{blue}{[ Liyi: #1 ]}}}
  \newcommand{\oth}[2]{\textbf{\textcolor{red}{[ #1: #2 ]}}}
  \newcommand{\xwu}[1]{\textbf{\textcolor{purple}{[ Xiaodi: #1 ]}}}
  
  \newcommand{\ynote}[1]{\textbf{\textcolor{magenta}{[ Yi: #1 ]}}}

  \colorlet{MZ}{violet!80!pink}

  \newcommand{\mzr}[1]{{\color{MZ}{#1}}}
  \newcommand{\was}[1]{}
  
\NewCommandCopy{\Creff}{\Cref}
  \renewcommand{\Cref}[1]{\mbox{\Creff{#1}}}

  \usepackage[inline]{enumitem}
  \colorlet{LC}{cyan!31!teal}

  \usepackage[inline]{enumitem}
}{
  \newcommand{\fixme}[1]{}
  \newcommand{\todo}[1]{}
  \newcommand{\rnr}[1]{}
  \newcommand{\mwh}[1]{}  
  \newcommand{\khh}[1]{}
  \newcommand{\liyi}[1]{}
  \newcommand{\shh}[1]{}
  \newcommand{\xwu}[1]{}
  \newcommand{\oth}[2]{}
  \newcommand{\mzr}[1]{}

  \newcommand{\ynote}[1]{}
}

\newtoggle{submission}
\toggletrue{submission}

\iftoggle{submission}{
  
}{
  
}

\makeatletter
\newcommand{\bigcupdot}{\mathop{\vphantom{\bigcup}\ooalign{$\m@th\bigcup$\cr \hidewidth$\m@th\cdot$\hidewidth\cr 
    }}\displaylimits
}
\makeatother \usepackage{pifont}\usepackage{multicol,tabularx,capt-of}
\usepackage{multirow}

\NewDocumentEnvironment{varsubfigure}{O{c}mo}
 {\begin{subfigure}[#1]{#2}
    \if#1t\vspace{0pt}\fi
    \makebox[0pt][r]{\refstepcounter{subfigure}\IfValueT{#3}{\label{#3}}(\thesubfigure) }\begin{tabular}{@{}p{\textwidth}@{}}
 }
 {\end{tabular}\if#1b\vspace{0pt}\fi\end{subfigure}}

\begin{document}
\title{DisQ: A Model of Distributed Quantum Processors (Extended Version)}                      

\titlerunning{DisQ: A Model of Distributed Quantum Processors}
\authorrunning{L. Chang et al.}

\author{Le Chang\inst{1} \and
Saitej Yavvari\inst{2} \and
Rance Cleaveland\inst{1} \and
Samik Basu\inst{2} \and
Runzhou Tao\inst{1} \and
Liyi Li\inst{2}}

\institute{University of Maryland, USA \\
\email{lchang21@umd.edu, rance@cs.umd.edu, rztao@umd.edu} \and
Iowa State University, USA \\
\email{saitej02@iastate.edu, sbasu@iastate.edu, liyili2@iastate.edu}}

\maketitle

\begin{abstract}
 The next generation of distributed quantum processors combines single-location quantum computing and quantum networking techniques to enable large entangled qubit groups to be established across remote processors, and for quantum algorithms to be executed distributively.
We present \disq, as the first formal model of distributed quantum processors, and permit the analysis of distributed quantum programs in the new computation environment.
The core of \disq is a distributed quantum programming language that combines the concepts of the Chemical Abstract Machine (CHAM) and Markov Decision Processes (MDP) to provide clearly distinguishable quantum concurrent and distributed behaviors. We also develop a simulation relation, based on classical simulation infrastructure, to verify the equivalence of a quantum algorithm and its distributed versions, enabling the equivalence check of the distributed version of a sequential quantum program.

 \end{abstract}

 \section{Introduction}
\label{sec:intro}

Quantum computing has shown significant promise for achieving quantum advantage by enabling the design of algorithms that are substantially faster than their classical counterparts. 
However, near-term intermediate-scale quantum (NISQ) devices face serious challenges in scaling up to execute practical quantum applications \cite{Caleffi:2022wxp,10488877}. A key limitation is that the quantum entanglement—a critical resource that powers many quantum algorithms—is constrained by the small size of current machines. For example, while implementing Shor’s algorithm would require approximately $5,000$ coherent and entangled qubits, existing single-location quantum computers can sustain only about $50$ such qubits.

\begin{wrapfigure}{l}{5cm}
  {\begin{center}
  \vspace*{-2em}
\includegraphics[width=0.35\textwidth]{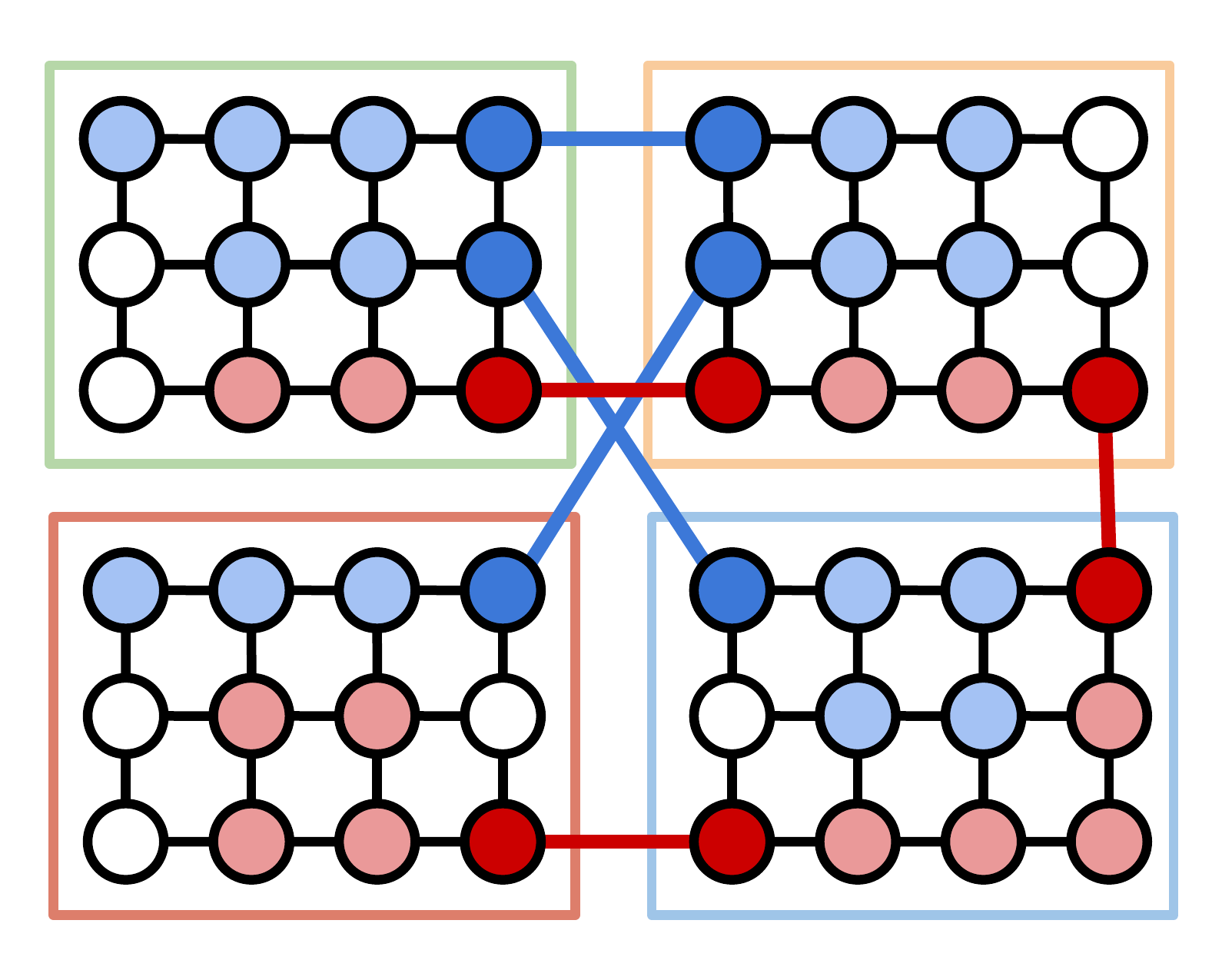}
\vspace*{-2em}
\end{center}}
   \caption{A distributed QPU structure with four QPUs, executing two programs (\textcolor{red}{red} and \textcolor{blue}{blue}). The thicker line communicates qubits from different QPUs. }
\label{fig:processor}
\end{wrapfigure}

To overcome the scalability limitations of current quantum computers, the next generation of quantum computing architectures is moving towards Distributed Quantum Computing (DQC), built on the foundation of interconnected Quantum Processing Units (QPUs) \cite{Chen2023,chu2024titandistributedlargescaletrappedion,main2024distributedquantumcomputingoptical,inc2024distributedquantumcomputingsilicon,ionqdistributed,googledis}, e.g, IonQ has announced plans to build a quantum computing system that will integrate over $1,000$ logical qubits through interconnected QPUs by 2027. As illustrated in \Cref{fig:processor}, a QPU typically consists of a local cluster of qubits (circle nodes in the figure) that can achieve high levels of entanglement, but the size of such entanglement remains limited by the hardware architecture. To interconnect multiple QPUs, photonic qubits and links (shown as darker nodes and thicker lines)\footnote{This approach uses photonic channels to distribute entanglement between QPUs.} enable entanglement sharing across different QPUs. These photonic qubits are often referred to as communication qubits, or "commq" for short. By linking QPUs in this way, quantum computers can create large-scale entangled states necessary to run complex quantum algorithms. Additionally, each QPU can also perform circuit parallelism to gain performance, i.e., a sequential quantum circuit is split into parallel components applied to different qubits.

Based on the above QPU scheme, it is necessary to develop distributed quantum programs with single-location parallelism. This faces three challenges: 1) we need a mechanism to develop a distributed quantum program with the ability to show its equivalence with respect to the sequential version, 2) such mechanism needs to consider the single-location parallelism, and 3) it is ideal to conduct such equivalence checking based on classical verification infrastructure \cite{MILNER19921}, which will be discussed below. 

We propose a programming language framework, named \disq, to model DQC systems and to use the model for developing distributed quantum programs, enabling equivalence checking between a sequential quantum program and its distributed version and allowing single-location quantum parallelism to be equated using a classical process algebraic framework. Our primary contribution is a faithful language model of DQC architectures. The emphasis is on managing quantum resources, such as qubit entanglement, across different QPUs and supporting the development of distributed quantum programs, acknowledging qubit resource limitations.

Previous works on quantum programming languages primarily focused on designing new systems for modeling and verifying quantum programs, e.g., quantum process algebras \cite{10.1145/1040305.1040318,10.1145/1507244.1507249,10.1145/2400676.2400680}, for modeling parallel quantum programs. Many of these systems define new kinds of operational semantics and equivalence relations tailored for quantum behaviors, such as entanglement and measurement. 

While these approaches are expressive, they often diverge from classical foundations and introduce abstractions, such as non-local gates or shared qubits across processes, thereby establishing a theoretical foundation for the quantum (bi)-simulation relation, where effective equivalence checking can be challenging. In addition, most of these models \cite{10.1145/1040305.1040318,10.1145/1507244.1507249} are designed for parallel or concurrent execution within a single quantum computer, rather than for systems that involve multiple quantum processors. In our work, we take a different approach. Instead of creating a completely new theory, we try to stay close to classical nondeterministic bi-simulation. We discuss the similarities between classical message passing and quantum remote communication, identify additional constraints of quantum communication, and enforce them by extending the language with a type system based on \emph{loci}, which describe the location of qubits and whether they may be entangled. This extension enables us to reuse classical simulation techniques to reason about correctness, ensuring that all quantum operations adhere to physical constraints.

Note that quantum distributed and parallel systems might exhibit not only nondeterministic behavior but also a probabilistic nature.
To model the quantum nature, we utilize classical probabilistic (bi)-simulation, viewing a \disq system as admitting a Markov decision process (MDP). This means that a \disq program exhibits both nondeterministic (across distributed QPUs) and probabilistic (quantum parallelism inside a QPU) behaviors.
To model this complication, \disq adopts \emph{membranes} from the Chemical Abstract Machine (CHAM) \cite{BERRY1992217} to represent different QPUs and impose different rules for inter-QPU distributed and intra-QPU parallel communications. Thus, all these quantum distributed and parallel behaviors can be reasoned about in a unified framework.

\disq aims to enable the development of distributed quantum programs based on sequential quantum programs, making the execution of non-trivial quantum programs possible in the near term through the following contributions.

\begin{itemize}
\item We introduce \disq, a core calculus with explicit locations, together with its syntax and a small-step operational semantics that \emph{separates} quantum operation probabilities from nondeterminism due to communication. An MDP view can be derived by installing a scheduler, enabling classical analysis without conflating the two phenomena. \footnote{The artifact is available at \url{https://github.com/lec9243/DisQ}}

\item We design a lightweight location-aware type system that captures quantum and distributed constraints (e.g., no-cloning and resource locality), ensuring well-formed communication and teleportation usage.

\item We develop an observational simulation to reason about equivalence between distributed programs—admitting both intra- and inter-membrane communication—and their sequential counterparts. In particular, we show that quantum teleportation \emph{refines} an abstract quantum channel.

\item We evaluate \disq on representative case studies, including distributed Shor's and quantum addition circuits, with additional details in the Appx. \ref{sec:morecases}.

\end{itemize}

 \section{Background} \label{sec:overview-background}

\noindent\textbf{\textit{Quantum Data and Computation.}}
A quantum datum (often called a quantum state) consists of one or more qubits. A single qubit is a normalized superposition $\ket{\psi}=z_1\ket{0}+z_2\ket{1}$ with $|z_1|^2 + |z_2|^2 = 1$. 
Multi-qubit data is formed by tensor products, but some joint states cannot be decomposed into separate single-qubit states; these are entangled states, e.g., the Bell pair $\frac{1}{\sqrt{2}}(\ket{00}+\ket{11})$.
Quantum computation evolves states by unitary gates, while measurement produces classical outcomes with probabilities determined by amplitudes and collapses the measured state. 
A common example is the Hadamard gate $\cn{H}$, which maps $\ket{0}$ to $\ket{+}=\frac{1}{\sqrt{2}}(\ket{0}+\ket{1})$ and $\ket{1}$ to $\ket{-}=\frac{1}{\sqrt{2}}(\ket{0}-\ket{1})$. Measuring $\ket{+}$ yields $\ket{0}$ or $\ket{1}$ with probability $\frac{1}{2}$ each. Superposition, entanglement, and probabilistic measurement are the quantum features most relevant to \disq.

The \textbf{no-cloning theorem} states that there is no physical operation that can copy an arbitrary unknown quantum state \cite{wootters1982single}. This matters in DQC as sending a qubit to another location cannot leave behind another usable copy of the same quantum information. In \disq, quantum communication is modeled as the relocation of ownership, which motivates one-shot channels and the locality constraints later enforced by the type system.

\noindent\textbf{\textit{Markov Chains and Decision Processes for Programming Semantics.}}
A Markov chain \cite{markov1906,markov1907} is a stochastic model describing a sequence of possible events in which the probability of each event depends only on the state attained in the previous event, and the probability of a program execution depends on the multiplication of the chain of probabilities of events.
It provides a standard, labeled transition description of defining the semantic behaviors of probabilistic programming by viewing probabilities as labels in semantic transitions; these labels are intrinsic and cannot be masked.
Markov decision process \cite{10.5555/528623} extends a Markov chain by combining a nondeterministic choice with a probabilistic transition. Here, every step of computation is essentially a combination of two steps: 1) a nondeterministic choice (the choice in \disq selects membrane locations for an event), and 2) a probabilistic move with a probability label.

\noindent\textbf{\textit{The CHAM Model}} (CHAM)~\cite{BERRY1992217} models distributed and concurrent behaviors as chemical reactions among molecules inside solutions. The concept of membranes allows processes to interact concurrently within a location, while airlocks enable controlled communication between different locations. This abstraction underpins our \disq language, where membranes represent local quantum computing units, and communication is modeled using explicit channels.
 \section{A Guided Example: Distributed Shor's Algorithm}
\label{sec:dshors}

\begin{wrapfigure}{r}{4cm}
\includegraphics[width=0.35\textwidth]{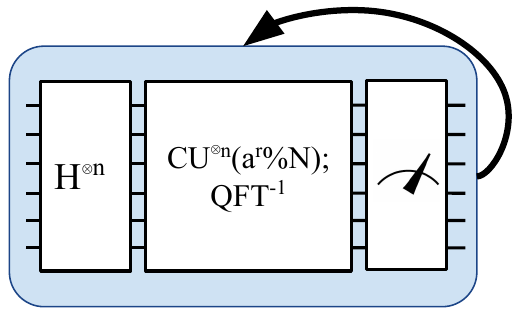}
 \vspace*{-0.5em}
   \caption{Shor's Flow}
 \vspace*{-1.5em}
 \label{fig:shors-full}
\end{wrapfigure}

To illustrate how \disq expresses and verifies distributed quantum programs, we present the development of the distributed version of Shor's algorithm, following a repeat-until-success pattern (\Cref{fig:shors-full}), where the classical post-processing may restart the quantum order-finding subroutine.
This example shows how locality, communication, and refinement interact in a distributed quantum algorithm.
Background information is given in \Cref{sec:overview-background}.

\begin{figure}[b]
 \vspace*{-1.5em}
{
\begin{minipage}[hb]{0.55\textwidth}
  \includegraphics[width=1\textwidth]{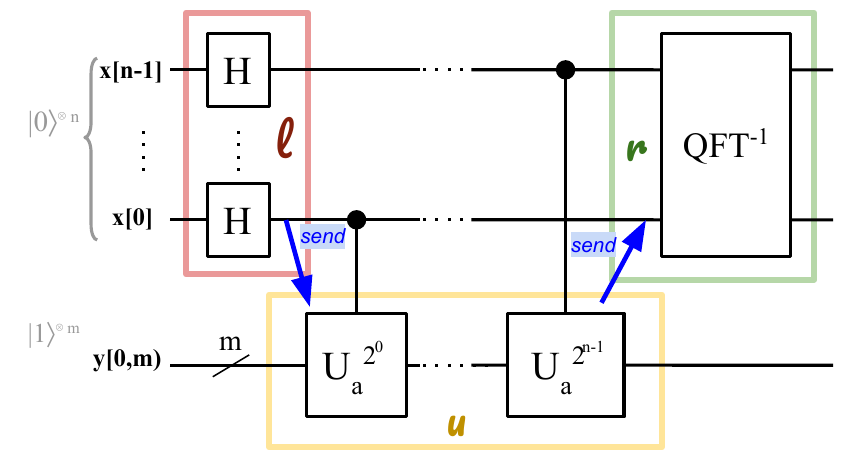}
  \vspace*{-1em}
       \caption{The order finding of Shor's algorithm, splitting into two pieces $l$, $u$, and $r$, executable in three QPUs, connected via photonic links.}
  \label{fig:shorseqa}
  \end{minipage}
    \hfill
    \begin{minipage}[hb]{0.42\textwidth}
  \includegraphics[width=1\textwidth]{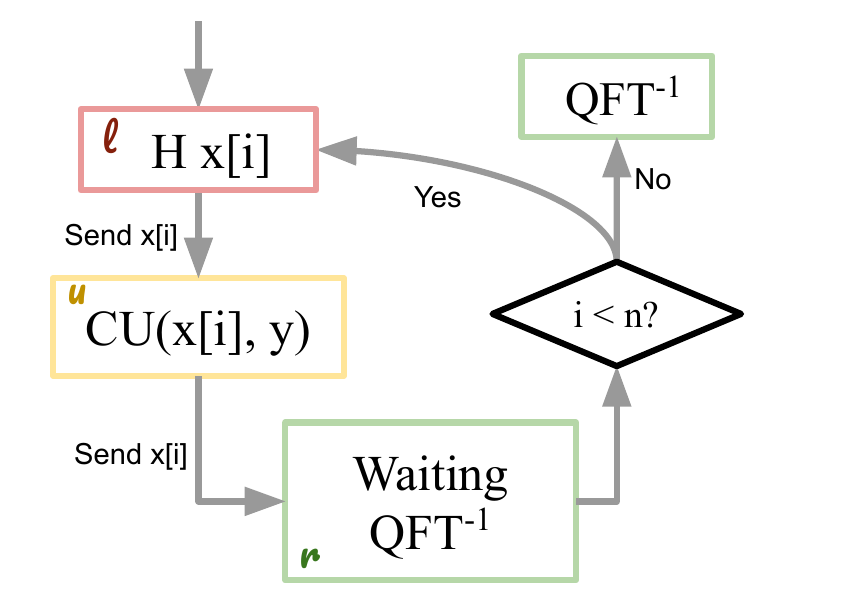}
  \vspace*{-1em}
 \caption{Distributed Transitions. A quantum channel $c(1)$ (state: $c[0]$) connects $l$ and $u$, and $c'(1)$ (state: $c'[0]$) connects $u$ and $r$.}
   \label{fig:shordisa}
  \end{minipage}
  }
   \vspace*{-1.5em}
\end{figure}

\myparagraph{Distributed Quantum Programs via Classical Process Algebra.}
The circuit in \Cref{fig:shorseqa} is a sequential Shor's algorithm quantum component (order finding) implementation, containing two registers $x$ and $y$, where $x$ represents the solution for the found order and $y$ stores the modular multiplication constraint. A practical order-finding implementation requires large registers (on the order of $n$ qubits for $x$ and about $2n$ for $y$ when factoring an $n$-bit $N$), quickly exceeding NISQ-scale devices.
The NISQ limitation necessitates more advanced quantum computing systems, and a distributed architecture is a possible solution.

Distributed quantum computation uses quantum networking techniques to connect multiple single-location QPUs; such communication can be described by quantum teleportation \cite{PhysRevLett.70.1895,Rigolin_2005}. 
To distribute the order-finding component, we split the circuit in \Cref{fig:shorseqa} into three parts, each executed on a separate QPU, and use quantum networking to communicate between them.

Before we discuss the details, we first demonstrate in \disq that such communication can be modeled by a classical messaging passing process algebra. Consider a simple grammar for communicating processes with the CHAM-based membranes in \disq (the full model is in \Cref{sec:disq}):

{\small
\vspace{-1em}
\[
\alpha ::= a \mid c(n)
\qquad
D ::= \downd{\alpha} \mid \csenda{\alpha}{v}\mid\creva{\alpha}{y} 
\qquad
U ::= U(\overline{x})
\]
\[
R ::= \zero \mid \seq{D}{R} \mid \seq{U}{R} \qquad
P ::= \scell{\overline{R}}_l \mid \sacell{R}{\overline{T}}_l
\]
}

Here, a process of type $R$ can be either a terminating process $\zero$ or a sequential process where its behavior evolves by either performing channel creation ($\downd{\alpha}$), a send-action ($\csenda{\alpha}{v}$: send $v$ over channel $a$), receive-action ($\creva{\alpha}{y}$: receive some data over channel $\alpha$ and write to $y$), or a quantum operation $U$. 
The membrane description $P$ is either a membrane $\scell{\ .\ }_l$ containing a multiset of processes of type $R$ denoted by $\overline{R}$ with explicit location information captured as $l$, or a membrane with an airlocked process $\sacell{R}{\overline{T}}_l$, where $R$ is ready to interact with some other airlocked process associated with a different membrane.
Intuitively, an airlocked process is a process temporarily isolated from its membrane so that it can synchronize with a matching process in another membrane.
A \disq program is a set of such membranes. Observe the inherent nondeterminism in the interactions between processes within each membrane and between processes across membranes. Any two processes in each membrane with appropriate sending/receiving actions may be non-deterministically selected for interaction; similarly, any two membranes with appropriate airlocked processes can be chosen for interactions across membranes.  This is similar to the CHAM model. 

Although classical and quantum communications differ, they exhibit similarities at an abstract level, where both can be described by the message-passing model presented above.
Classically, two processes interact (synchronize) by sending and receiving messages over the same channel.
Quantumly, if we model a quantum teleportation (\Cref{sec:msgpassing}) as a quantum channel, a quantum message is relocated between two processes via the channel.
Consider the following \disq program between two membranes $l$ and $r$, where the left side is the program transition and the right side is the quantum state the program is applied to.

{\footnotesize
\[
\begin{array}{lcl@{\qquad}l}
\textcolor{blue}{(1)}
&&\sdot{\textcolor{orange}{\bscell{\sdot{\seq{\downd{\alpha}}{\seq{\csenda{\alpha}{\theta_1}}}{\zero}}{\zero}}_l}}{\textcolor{purple}{\bscell{\sdot{\seq{\downd{\alpha}}{\seq{\creva{\alpha}{y}}{\zero}}}{\zero}}_r}}
&
\textcolor{spec}{\bglocus{\theta_1\sqcupplus \theta_2}{l}: \Phi}
\\
\textcolor{blue}{(2)}
&
\xrightarrow{l.\frac{1}{2}}
&\sdot{\textcolor{orange}{\bsacell{{\seq{\downd{\alpha}}{\seq{\csenda{\alpha}{\theta_1}}}{\zero}}}{\zero}_l}}{\textcolor{purple}{\bscell{\sdot{\seq{\downd{\alpha}}{\seq{\creva{\alpha}{y}}{\zero}}}{\zero}}_r}}
&
\textcolor{spec}{\bglocus{\theta_1\sqcupplus \theta_2}{l}: \Phi}
\\
\textcolor{blue}{(3)}
&
\xrightarrow{r.\frac{1}{2}}
&\sdot{\textcolor{orange}{\bsacell{{\seq{\downd{\alpha}}{\seq{\csenda{\alpha}{\theta_1}}}{\zero}}}{\zero}_l}}{\textcolor{purple}{\bsacell{{\seq{\downd{\alpha}}{\seq{\creva{\alpha}{y}}{\zero}}}}{\zero}_r}}
&
\textcolor{spec}{\bglocus{\theta_1\sqcupplus \theta_2}{l}: \Phi}
\\
\textcolor{blue}{(4)}
&
\xrightarrow{l.r.1}
&\sdot{\textcolor{orange}{\bscell{\sdot{\seq{\csenda{\alpha}{\theta_1}}{\zero}}{\zero}}_l}}{\textcolor{purple}{\bscell{\sdot{\seq{\creva{\alpha}{y}}{\zero}}{\zero}}_r}}
&
\textcolor{spec}{\bglocus{\theta_1\sqcupplus \theta_2}{l}: \Phi}
\\
\textcolor{blue}{(5)}
&
\xrightarrow{l.\frac{1}{2}}
&\sdot{\textcolor{orange}{\bsacell{\seq{\csenda{\alpha}{\theta_1}}{\zero}}{\zero}_l}}{\textcolor{purple}{\bscell{\sdot{\seq{\creva{\alpha}{y}}{\zero}}{\zero}}_r}}
&
\textcolor{spec}{\bglocus{\theta_1\sqcupplus \theta_2}{l}: \Phi}
\\
\textcolor{blue}{(6)}
&
\xrightarrow{r.\frac{1}{2}}
&\sdot{\textcolor{orange}{\bsacell{\seq{\csenda{\alpha}{\theta_1}}{\zero}}{\zero}_l}}{\textcolor{purple}{\bsacell{\seq{\creva{\alpha}{y}}{\zero}}{\zero}_r}}
&
\textcolor{spec}{\bglocus{\theta_1\sqcupplus \theta_2}{l}: \Phi}
\\
\textcolor{blue}{(7)}
&
\xrightarrow{l.r.1}
&\sdot{\textcolor{orange}{\bscell{\sdot{\zero}{\zero}}_l}}{\textcolor{purple}{\bscell{\sdot{\zero}{\zero}}_r}}
&
\textcolor{spec}{\glocus{\theta_2}{l} \sqcupplus \glocus{\theta_1}{r}: \Phi}
\\
\textcolor{blue}{(8)}
&
\xrightarrow{l.1}
&{\textcolor{orange}{\bscell{\sdot{\zero}{\zero}}_r}}
\quad
\xrightarrow{r.1}
\quad
{\textcolor{purple}{\emptyset}}
&
\textcolor{spec}{\glocus{\theta_2}{l} \sqcupplus \glocus{\theta_1}{r}: \Phi}
\end{array}
\]
}

Lines (1) to (4) create a quantum channel between $l$ and $r$.
The transitions in (2) and (3) select a process inside membranes $l$ and $r$, respectively, with a $\frac{1}{2}$ probability,
i.e., each membrane contains two processes, and the probability of selecting any one is half.
The transition in line (4) creates the quantum channel $\alpha$ (we do not show its state here for simplicity).
\disq uses a process-algebraic style of name scoping to ensure that every quantum channel is created before it is used.
The transitions in lines (5) to (7) utilize the quantum channel $\alpha$ to convert a quantum message $\theta_1$ from the membrane $l$ to $r$,
where lines (5) and (6) respectively select the two processes in the membranes $l$ and $r$, and line (7) performs a traditional message communication from $l$ to $r$.

The effect of message communication occurs in the quantum state, where it transforms a quantum message between two membranes, mimicking quantum teleportation.
\disq utilizes a locus structure $\textcolor{spec}{\glocus{\theta_1 \sqcupplus \theta_2}{l}}$ to indicate that two qubit arrays $\theta_1$ and $\theta_2$ are entangled and locate in the membrane $l$, indicated by the operation $\textcolor{spec}{\sqcupplus}$.
After the communication transition in line (7), $\theta_1$ is transformed to the membrane $r$.
The physical meaning of the transformation is that we destroy the qubit array $\theta_1$ in the membrane $l$ and reproduce it in $r$, while preserving all its information, using the same name $\theta_1$ to indicate the phenomenon.
Note that quantum entanglement is also a piece of information to be preserved; therefore, after the transformation, $\theta_1$ is still entangled with $\theta_2$ in $l$. We use the structure $\textcolor{spec}{\glocus{\theta_2}{l} \sqcupplus \glocus{\theta_1}{r}}$ to indicate that $\theta_1$ and $\theta_2$ are entangled, but they are from two different membranes.

To develop the distributed order finding algorithm, we aim to place the operations for the $x$ and $y$ registers on different QPUs. Furthermore, each qubit in the $x$ registers is applied by an individual Hadamard operation without any connection, indicating that these qubits can perform gate operations in sequence to ensure that we only manipulate one qubit in the $l$ location at a time. Below, we distribute the order-finding operation across three membranes.

\begin{example}[One Step Distributed Shor's Algorithm]\label{def:shors}\rm
We show the three membranes, $l$, $u$, and $r$, below for the computation in \Cref{fig:shorseqa},
assuming having a $1$-qubit quantum channel $c(1)$ between $l$ and $u$, and $c'(1)$ between $u$ and $r$.

{\scriptsize
\begin{center}
$
\sdot{\sdot{\bscell{\seq{\ssassign{x[i]}{}{\cn{H}}}{\seq{\csenda{c(1)}{x[i]}}{\zero}}}_l}
{\bscell{\seq{\creva{c(1)}{w}}{\seq{\ssassign{w\sqcupplus y[0,n)}{}{\cn{CU}(v^{2^i})}}{\seq{\csenda{c'(1)}{w}}}{\zero}}}_u}}
{\bscell{{\seq{\creva{c'(1)}{q}}{...}}}_r}
$
\end{center} 
}
\end{example}

Assume that $x$ and $y$ registers have $n$ qubits and $x[i]$ refers to the $i$-th qubit in $x$ and $y[0,n)$ to refer to the range $[0,n)$ of the qubit array $y$.
We restructure the operations in membranes $l$ and $u$ (\Cref{fig:shorseqa,fig:shordisa}), to be a repetition of two operations ${\cn{H}(x[i])}$ and ${\cn{CU}(v^{2^i})(x[i],y[0,n))}$ for $i\in[0,n)$ (\disq uses the program syntax format $\ssassign{x[i]}{}{\cn{H}}$ and $\ssassign{x[i]\sqcupplus y[0,n)}{}{\cn{CU}(v^{2^i})}$).
Each single step applies a Hadamard gate to $x[i]$ in $l$ and transmit $x[i]$'s quantum information to $u$, via the channel $c(1)$, so that a control gate ($\ssassign{x[i]\sqcupplus y[0,n)}{}{\cn{CU}(v^{2^i})}$) can be applied to. The operation controls on $x[i]$ and applies a modulo-multiplication to $y[0,n)$.
We then utilize the $c'(1)$ channel to transmit $x[i]$ again to the membrane $r$.

\myparagraph{Enforce Quantum Constraints.}
Quantum communication introduces additional restrictions due to the no-cloning principle, which imposes constraints on the system.
For example, after the channel $c(1)$ and $x[i]$ are used in the send operation in the membrane $l$, one cannot apply additional operations to them in $l$ anymore.
We discover the following constraints.

\begin{definition}[Quantum Message and Channel Well-formed Constraints]\label{def:chanconstrant}\rm
We define the constraint below for a correct \disq program \emph{in a membrane}.

\begin{enumerate}
\item Every quantum channel $c$ must be initialized before being used, and used only once as a quantum message passing channel.
\item A quantum message $\theta$ cannot appear in a later execution after it is sent.
\item A quantum operation $U(\overline{x})$ cannot have overlapping arguments in $\overline{x}$ and $U(\overline{x})$'s function body guarantees no-cloning.
\end{enumerate}
\end{definition}

The constraint (1) ensures that quantum channels are always one-time, while (2) ensures that a quantum message cannot be cloned but relocated.
The constraint (3) ensures the no-cloning property in a sequential quantum operation.
To ensure the constraints, \disq utilizes a locus structure, a qubit collection indicating entangled groups, and ensures that qubits in two loci are not entangled.
Before the order finding algorithm executes, we represent $x$ and $y$ registers as two separate loci $\cglocus{x[0,n)}{l}$ and $\cglocus{y[0,n)}{r}$, indicating that they are located in membrane $l$ and are not entangled. After a single application on $x[0]$ in \Cref{def:shors}, we connect $x[0]$ with $y$ registers to be a locus $\cglocus{x[0]}{r}\, \textcolor{spec}{\sqcupplus}\, \cglocus{y[0,n)}{r}$, indicating that $x[0]$ is relocated in membrane $r$ and joined the entangled group $\cglocus{y[0,n)}{r}$, while the other locus becomes $\cglocus{x[1,n)}{l}$. \disq uses loci to characterize quantum resources in different membranes via typing, which is then used to partition the quantum state when evaluating programs, in order to guarantee \Cref{def:chanconstrant}. 

\begin{wrapfigure}{r}{4cm}
\vspace*{-1em}
  \includegraphics[width=0.3\textwidth]{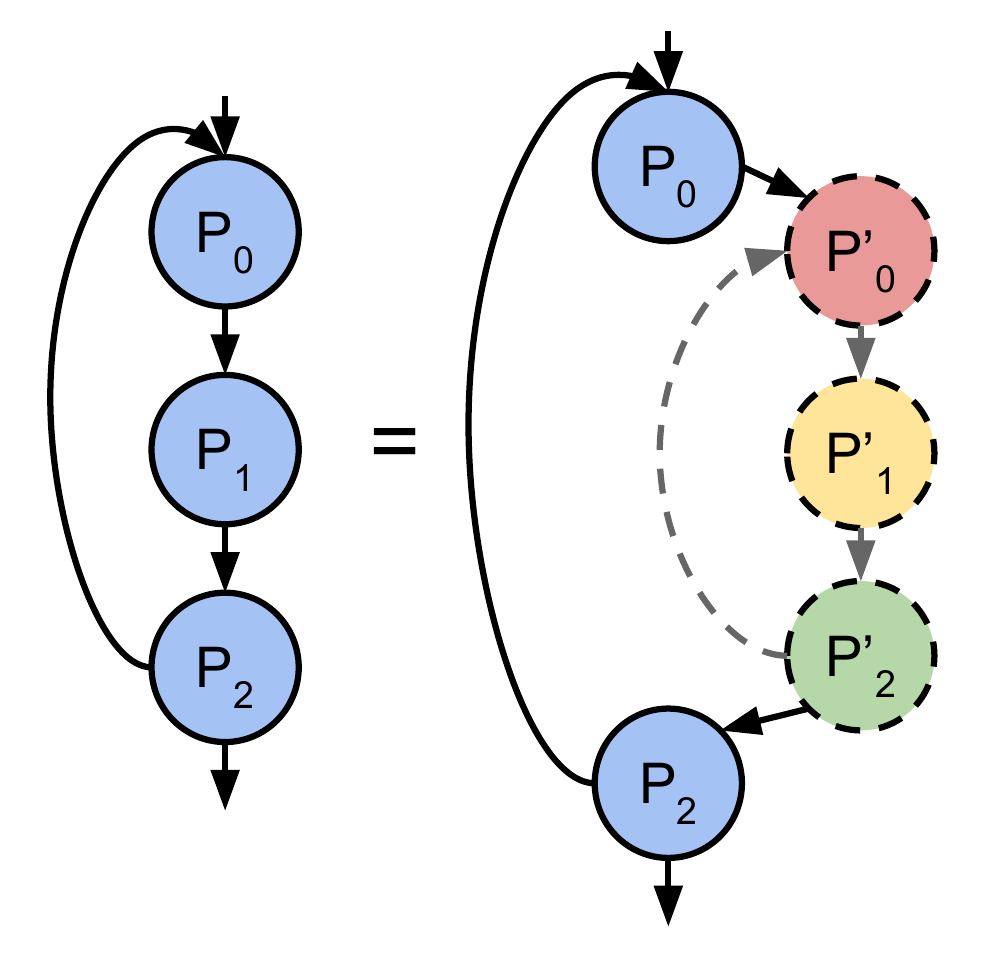}
\vspace*{-0.5em}
  \caption{Simulation diagram for the sequential and distributed order-finding programs. The intermediate communication steps in the distributed version are treated as $\tau$-steps, so the simulation compares the observable checkpoints, such as $p_0$ and $p_2$.}
\vspace*{-1.5em}
\label{fig:aequiv-ex}
\end{wrapfigure} 

\myparagraph{Equivalence Check of the Distributed and Sequential Programs via Classical Simulation.}\label{sec:equivcheck}
One of our contributions is the demonstration of using classical simulation to equate a sequential program and its distributed version.
To simulate the distributed and sequential programs in \Cref{fig:shorseqa,fig:shordisa}, we can use the traditional weak simulation relation \cite{MilnerCCS} to judge the equivalence between the sequential (named as $P$) and distributed versions (named as $Q$) of Shor's algorithm, by equating the resulting quantum states of executing the two programs, i.e., the simulation relation $(\Phi,P) \sim (\Phi,Q)$ can be defined as, for every $(\Phi,P) \to (\Phi', P')$, if the transition is not a $\tau$ step, we can find an equivalent transition $(\Phi, Q) \to (\Phi', Q')$ and $(\Phi',P') \sim (\Phi',Q')$.
\Cref{fig:aequiv-ex} demonstrates the procedure of constructing the simulation relation for the two order finding versions (\Cref{fig:shorseqa,fig:shordisa}).
Here, we recognize the $\tau$ steps to be message send and receive operations in the distributed version.

\myparagraph{Extend Classical Simulation Including Probabilistic Features.}\label{sec:probnature}
The above simulation is a special case because the order-finding fragment contains no measurements, and each membrane has only one active process in our one-step presentation.
In general, \disq executions exhibit probabilistic behaviors from two sources.
First, quantum measurements induce probabilistic branching, and the full Shor workflow (\Cref{fig:shors-full}) repeats the quantum subroutine depending on classical post-processing outcomes.
Second, even without measurements, intra-membrane interleavings introduce probabilistic choices when a membrane contains multiple eligible processes, since the operational semantics selects one process to progress at each step.
To accommodate both aspects, \disq views executions as Markov decision processes (MDPs): nondeterminism models scheduling/interleavings across membranes, while probabilities arise from measurements and local process selection.
Our observable simulation (\Cref{sec:disq-equiv}) compares programs by grouping internal communication/synchronization steps as $\tau$-moves and by matching the probability mass of reaching corresponding configurations at a set of user-defined synchronization points.
This design enables equivalence/refinement checking between a sequential specification and its distributed realization in the presence of both communication and probabilistic behaviors.
 
\section{\disq Formalism}
\label{sec:disq}

This section presents the \disq's state representation, syntax, semantics, type systems, and metatheorems.

\subsection{\disq State Representation}\label{sec:state}

\begin{figure}[t]
{\scriptsize
\[
\begin{array}{l}
\textcolor{blue}{\text{Basic Terms:}}\\
{\hspace*{-0.3em}
\begin{array}{l@{\;\;}l@{\;\;}c@{\;\;}l @{\quad} l@{\;\;}l@{\;\;}c@{\;\;}l @{\quad} ll@{\;\;}c@{\;\;}l @{\quad} l@{\;\;}l@{\;\;}c@{\;\;}l @{\qquad} l @{\;\;} l @{\;\;}c@{\;\;}l}
      \text{Amplitude} & z & \in & \mathbb{C}
&
\text{Nat} & i,j,m, n & \in & \mathbb{N}
&
      \text{Variable/C-Chan} & x,y,a &&
\\
      \text{Bit} & b & ::= & 0\mid 1  
 &
      \text{Basis Vector}& \beta & ::= & (\ket{d})^*

&
  
  \text{Q-Chan Name}& c &&    
        \\
   \text{Bitstring} & d & \in & b^{*}     
   & 
      \text{Location}& l,r,u &&
\end{array}
}
\\[0.7em]
\textcolor{blue}{\text{Kinds and Classical/Quantum Data:}}\\
\begin{array}{llcl@{~}c@{\quad}l} 
      \text{Kind} & g & ::= & \cmode  &\mid& \qmode{n}\\
      \text{Classical Scalar Data} & v & ::= & d &\mid & \theta\\
      \text{Basic Ket} & \eta & ::= & z \beta\\
      \text{Quantum Data} & \theta & ::= & \ssum{j=0}{m}{\eta_j}
    \end{array}
\\
\textcolor{blue}{\text{Quantum Loci, Environment, and States}}\\
\begin{array}{llclcl} 
      \text{Qubit Array Range} & s & ::= & x\srange{n}{m} \\
      \text{Local Locus} & \kappa & ::= & \overline{s}  & \text{concatenated op} & \sqcupplus \\
      \text{Locus} & K & ::= & \overline{\glocus{{\kappa}}{l}}  & \text{concatenated op} & \sqcupplus \\
\text{Local Quantum State (Heap)} & \varphi & ::= & \overline{ \kappa : \theta } & \text{concatenated op} & \cupdot \\
      \text{Quantum State (Heap)} & \Phi & ::= & \overline{ K : \theta } & \text{concatenated op} & \cupdot 
    \end{array}
\\[0.7em]
\textcolor{blue}{\text{Syntax Abbreviations and Basis/Locus Equations}}\\
{\scriptsize
  \begin{array}{r @{~} c @{~} l @{\qquad}  r@{~}c @{~} l @{\qquad}  r @{~}c @{~} l @{\qquad} r @ {~} c @{~} l}
    1 \beta &\equiv& \beta
&
   \sum_{j=0}^{0}\eta_j &\equiv& \eta_0
&
x[n,n\splus 1) &\equiv& x[n]
&
\glocus{\kappa\sqcupplus \kappa'}{l} &\equiv& \glocus{\kappa}{l}\sqcupplus\glocus{\kappa'}{l}
\\[0.2em]
\emptyset \sqcupplus \kappa & \equiv & \kappa
&
x[n,n) & \equiv & \emptyset
&
\ket{d_1}\ket{d_2} &\equiv& \ket{d_1 d_2}
&
x[n,m) & \equiv & x[n,j)\sqcupplus x[j,m) \,\;\cn{if}\;j \in [n,m]
    \end{array}
}
\end{array}
  \]
}  
\vspace*{-1em}
  \caption{\disq data element. Each range $x\srange{n}{m}$ represents the number range $[n,m)$ in the physical qubit array $x$. The operations after "concatenated op" are the concatenations for loci and quantum program states. Term $a$ is no more than a variable, but also represents classical channels. $c$ is a quantum channel name.}
\vspace*{-1em}
\label{fig:disq-state}
\end{figure}

We first discuss our quantum state representation in \Cref{fig:disq-state}.
There are two kinds of data: scalar ($\cmode$) and quantum ($\qmode{n}$, representing $n$ qubit arrays). 
For simplicity, variables and locations are in distinct categories. 
Quantum data valuations are represented using a varied Dirac notation $\sch{m}{z_j}{\beta_j}$, where $m$ is the number of basis-kets in the quantum data; each basis-ket contains an amplitude $z_j$ and a basis vector $\beta_j$, with $\slen{\beta_j}=n$ for all $j$, if the datum has kind $\qmode{n}$, meaning that the datum represents a $n$-qubit quantum array.

Quantum data are conceptually stored as a heap (a quantum state $\Phi\triangleq\overline{K : \theta}$), partitioned into regions described as \emph{loci} ($K$) in \disq; each region contains possibly entangled qubits, with the guarantee that cross-locus qubits are not entangled.
We use loci to group possibly entangled qubits and ensure qubit disjointness; each locus can be viewed as a chain of disjoint region segments labeled with explicit information about the location of local state variables, e.g.,  $\textcolor{spec}{\glocus{c[0]}{l}\sqcupplus \glocus{c[0]}{r}}$ suggests that the two qubits, both named $c[0]$, in locations $l$ and $r$ are possibly entangled. Note that the $\textcolor{spec}{}{l}$ notation in loci captures location information.
In describing a local quantum state ($\varphi$) in a membrane, we disregard the location information;
we can utilize local loci ($\kappa$) to refer to a quantum datum locally to a specific location.
Each local locus consists a list of \emph{disjoint ranges} ($s$), each represented by \(x\srange{n}{m}\)---an in-place array slice selected from \(n\) to \(m\) (exclusive) in a physical qubit array \(x\) (always being $\qmodename$ kind).
Ranges in a local locus are pairwise disjoint, written as \(s_1 \sqcupplus s_2\).

Each element in a quantum state $\Phi$ maps a locus $K$ to a quantum datum $\textcolor{spec}{\ssum{j=0}{m}{z_j\ket{d_j}}}$, with $\slen{K} \le \slen{d_j}$, i.e., the qubit length in $K$ might be less than the bitstring length in a basis-vector $\ket{d_j}$.
Essentially, a locus $K$ acts as a sequence of pointers pointing to entangled qubits, with information partly stored in $\ket{d_j}$.
We call the corresponding basis vector bits of qubits or locus fragments for a datum (or a basis-ket set) as the \textit{qubit's/locus's position bases} of the datum (or the basis-ket set).
In analyzing a local program piece, one might refer to part of the locus but we cannot simply separate out an entangled qubit state because it is not separable.
In performing such locality analysis, we shrink the locus $K$ but leave the quantum datum unchanged, and refer to the basis-vector locations $t \in [\slen{K}, \slen{d_j})$ as the locus $K$'s \emph{unreachable basis-vectors}.

\begin{example}[Unreachable Basis-vector Example]\label{def:example3}\rm

{\scriptsize
  \begin{mathpar}
   \inferrule[]
   { (\textcolor{spec}{\{ x[0] : \sum_{j=0}^1 \frac{1}{2} \ket{j}}\textcolor{stack}{\cmsg{\ket{j}}}\textcolor{spec}{\}} ,\seq{\qass{x[0]}{\cn{X}}}{\zero})
     \xrightarrow{1} (\textcolor{spec}{\{x[0]  : \sum_{j=0}^1 \frac{1}{2}\ket{\neg j}}\textcolor{stack}{\cmsg{\ket{j}}}\textcolor{spec}{\}} ,\zero)  }
   { (\textcolor{spec}{\{\glocus{x[0]}{l}\sqcupplus \glocus{x[0]}{r}:\sum_{j=0}^1\frac{1}{\sqrt{2}}\ket{j}\ket{j}\}}
    ,\scell{\seq{\qass{x[0]}{\cn{X}}}{\zero}}_r) \xrightarrow{r.1} 
    (\textcolor{spec}{\{\glocus{x[0]}{l}\sqcupplus \glocus{x[0]}{r}:\sum_{j=0}^1\frac{1}{\sqrt{2}}\ket{\neg j}\ket{j}\}}
    ,\scell{\zero}_r) }
  \end{mathpar}
}
\end{example}

The example shows locality and unreachable basis vectors.
$\textcolor{spec}{\glocus{x[0]}{l}\sqcupplus \glocus{x[0]}{r}}$ contains qubits in membranes $l$ and $r$.
When applying $\cn{X}$ gate to the locus locally in $r$, we first localize the locus to focus on $x[0]$ in $r$.
In defining the operation semantics inside the membrane $r$, we want to focus locally in $r$ and push the part in $l$ to unreachable positions, as the local quantum state $\textcolor{spec}{x[0] : \sum_{j=0}^1 \frac{1}{2} \ket{j}}\textcolor{stack}{\cmsg{\ket{j}}}$ appearing on the top.
Here, the first $\textcolor{spec}{\ket{j}}$ is $x[0]$'s position basis, and the second one is the unreachable basis vectors.
In this paper, we \textcolor{stack}{color} them with a hat. 

The bottom of \Cref{fig:disq-state} presents some notational convenience and syntactic equivalences ($\equiv$), e.g., we abbreviate a singleton range $x\srange{j}{j\splus 1}$ as $x[j]$.

\subsection{Syntax for The \disq Language}
\label{sec:disqsyntax}

\begin{figure}[t]
{
  \scriptsize
  \[\begin{array}{l@{\quad}l@{\;\;}c@{\;\;}l} 
    \text{Unitary Expr} & \mu\\
    \text{Bool Expr} & B \\
      \text{Channels} & \alpha & ::= & a \mid c(n) \\
      \text{Local Action} & U & ::= & \downd{x(n)} \mid \ssassign{\kappa}{}{\mu} \mid \smea{x}{\kappa} \\[0.2em]

      \text{Communication Action} & D & ::= &\downd{c(n)} \mid  \csenda{\alpha}{v}\mid\creva{\alpha}{y} \\[0.2em]
      
      \text{Process} & R,T,M,N & ::= & \zero \mid \seq{D}{R} \mid \seq{U}{R}\mid \sifb{B}{R}{T} \\[0.2em]
 
      \text{Membrane} & P,Q &::=& \scell{\overline{R}}_l \mid \sacell{R}{\overline{T}}_l
    \end{array}
  \]
}
\vspace*{-1em}
  \caption{\disq Syntax. Syntactic sugar: $\seq{\sifc{B}{R}}{T}=\sifb{B}{\seq{R}{T}}{T}$.}
  \label{fig:vqimp}
\vspace*{-1em}
\end{figure}

A \disq program is described as a multiset of location-specific quantum processes, with syntax (Figure~\ref{fig:vqimp}) over a membrane. There are two types of membranes: a multiset of processes ($\scell{\overline{R}}_l$) with location information ($l$) and an airlocked process associated with a membrane ($\sacell{R}{\overline{T}}_l$).
A process $R$, localized to a membrane, can be understood as a sequence of local quantum ($U$) or communication actions ($D$). We permit process algebraic message transmission operations ($D$ typed, $\csenda{\alpha}{v}$ and $\creva{\alpha}{y}$); they are the only communication actions that can perform direct message passing between different membranes, which can perform both classical and quantum message passing, depending on if the channels are classical or quantum.
The operation  $\downc{c}{n}$ creates a quantum channel ($c$).
If we have the process $\sacell{\seq{\downc{c}{n}}{R_1}}{\overline{T}_1}_l, \sacell{\seq{\downc{c}{n}}{R_2}}{\overline{T}_2}_r$, interacting $l$ and $r$,
which collaboratively create an $2n$-qubit Bell pair, each membrane shares an $n$ qubit array, pointed to by the locus $\textcolor{spec}{\glocus{c[0,n)}{l}\sqcupplus \glocus{c[0,n)}{r}}$. This is similar to $\pi$-calculus style creation of new channels. 

A local quantum operation can be a new blank ($\ket{0}$) qubit array generator $\downd{x(n)}$, a unitary operation $\ssassign{\kappa}{}{\mu}$, applying a unitary operation $\mu$ to a local locus $\kappa$, and quantum measurement $\smea{x}{\kappa}$, measuring the qubits referred to by $\kappa$ and storing the result as $x$. In \disq, we abstract away the detailed implementation of $\mu$ and assume that they can be analyzed by some systems describing quantum unitary circuits, e.g., VOQC \cite{VOQC}. We also permit a classical conditional $\sifb{B}{R}{T}$. The expression $B$ is an arbitrary classical Boolean expression, implemented using bit-arithmetic, i.e., $1$ as $\cn{true}$ and $0$ as $\cn{false}$.

\subsection{\disq Semantics}\label{sec:semantics}

\disq models the probabilistic aspect of quantum computation via probabilistic transitions, enabling single-location parallelism, based on a combination of Markov chains and Markov decision processes, and is divisible into two categories: process-level and membrane-level semantics.
The process level semantics is shown in \Cref{fig:exp-semantics-1}, as a labeled transition system $(\varphi, R)\xrightarrow{p}(\varphi', T)$, where $R$ and $T$ are processes, $\varphi$ and $\varphi'$ are the pre- and post- local quantum states described in \Cref{fig:disq-state}, and $p$ is the probability of the single step transition.
The membrane level semantics defines the nondeterministic behaviors of a \disq program, shown in \Cref{fig:exp-semantics-2}, formalized as a labeled transition system $(\Phi,\overline{P})\xrightarrow{\xi.p}(\Phi',\overline{Q})$ where $\xi$ (either  $l$ or $l.r$) captures the membrane locations ($l$ or $l.r$) participating in the nondeterministic choice of the transition, $p$ is the transition probability, and $\Phi$ and $\Phi'$ are the global pre- and post- quantum states in \Cref{fig:disq-state}.
We show selected rules, while the remaining are in \Cref{sec:semanticsa}.

\begin{figure*}[t]
{\scriptsize
  \begin{mathpar}        
      \inferrule[S-OP]{}{ 
(\varphi\cupdot \textcolor{spec}{\{\kappa \sqcupplus\kappa':  \theta\}},\ssassign{\kappa}{}{\mu}.R) \xrightarrow{1} (\varphi\cupdot \textcolor{spec}{\{\kappa\sqcupplus\kappa': \denote{\mu}^{\slen{\kappa}}\theta\}},R) }

       \inferrule[S-Self]{}{ (\varphi,\zero) \xrightarrow{1} (\varphi,\zero)}
               
    \inferrule[S-Mea]{p= \sum_j \slen{z_j}^2}{(\varphi\cupdot \textcolor{spec}{\{\kappa\sqcupplus\kappa': \sum_j z_j\ket{d}{\ket{d_j}}+\qfun{\theta}{\kappa,d \neq \kappa}\}},\smea{x}{\kappa}.R) \xrightarrow{p} (\varphi\cupdot \textcolor{spec}{\{\kappa' :   \sum_j {\frac{z_j}{\sqrt{p}}}{\ket{d_j}}\}}, {R[d/x]}) }
  \end{mathpar}
}
{\footnotesize
\begin{center}
$
\begin{array}{lclcl}
\denote{\mu}^{n}(\sum_{j}{z_j\ket{d_j}\beta_j})
&\triangleq&
 \Msum_{j}{z_j(\denote{\mu}\ket{d_j})\beta_j} & \texttt{where}& \forall j\,\slen{d_j}=n
 \\[0.2em]
 \qfun{(\Msum_{i}{z_i}{\ket{d_{i}}}{\beta_i}+\theta)}{\kappa,b} &\triangleq& \Msum_{i}{z_i}{\ket{d_{i}}}{\beta_i}
      &\texttt{where}&\forall i.\,\slen{d_{i}}=\slen{\kappa}\wedge \denote{b[d_{i}/\kappa]}=\texttt{true}
\end{array}
$
\end{center}
}
\vspace{-0.8em}
\caption{Selected \disq single process semantic rules.}
\label{fig:exp-semantics-1}
\vspace{-1em}
\end{figure*}

\noindent\textbf{\textit{Process Level Semantics.}}
A \disq process is a sequence of actions, and the rules in \Cref{fig:exp-semantics-1} define the selected semantic rules for a local action prefixed in a process.
Rule \rulelab{S-Self} shows that the process semantics in a membrane is reflexive and can make a move to itself to preserve the stochastic property in a Markov chain, explained shortly below.
Rule \rulelab{S-OP} applies a quantum unitary operation to a locus $\kappa$'s quantum data.
Here, the locus fragment $\kappa$ to which the operation is applied must be prefixed in the locus \(\kappa \sqcupplus \kappa'\) that refers to the entire quantum data \(\theta\).
If not, we will first apply equivalence rewrites, explained in \Cref{sec:typing} and \cite{li2024qafny}, to move $\kappa$ to the front.
With \(\kappa\) preceding the rest fragment \(\kappa'\), the operation's semantic function $\denote{\mu}^{n}$ is then applied to $\kappa$'s position bases in the quantum value \(\theta\).
More specifically, the function is only applied to the first $n$ (equal to $\slen{\kappa}$) basis bits of each basis-ket in the value while leaving the rest unchanged.
In \Cref{def:example3}, to apply $\cn{X}$ to the fragment $\cglocus{x[0]}{r}$ of $\textcolor{spec}{\glocus{x[0]}{l}\sqcupplus \glocus{x[0]}{r}}$, we use equivalence rewrites to ensure that $\cglocus{x[0]}{r}$ is prefixed in the locus and it is arranged as $\cglocus{x[0]}{r}$ followed by $\cglocus{x[0]}{l}$.
A measurement ($\smea{x}{\kappa}.R$) collapses qubits in a locus $\kappa$, binds a $\cmode$-kind integer to $x$, and restricts its usage in $R$.
Rule \rulelab{S-Mea} shows the partial measurement behavior.
Assume that the locus is $\kappa \sqcupplus\kappa'$; the measurement is essentially a two-step array filter:
(1) the basis-kets of the value is partitioned into two sets (separated by $+$): $\textcolor{spec}{(\scha{m}{z_j}{d}{\ket{d_{j}}})+\qfun{\theta}{\kappa,d\neq \kappa}}$, by randomly picking a $\slen{\kappa}$-length basis $d$ where every basis-ket in the first set have $\kappa$'s position basis $d$; and (2) we create a new array value by removing all the basis-kets not having $d$ as prefixes (the $\qfun{\theta}{\kappa,d\neq \kappa}$ part) and also removing the $\kappa$'s position basis in every remaining basis-ket; thus, the quantum value becomes $\textcolor{spec}{\sch{m}{\frac{z_j}{\sqrt{p}}}{\ket{d_{j}}}}$. 
Since the amplitudes of basis-kets must satisfy $\Msum_i\slen{z_i}^2=1$, we need to normalize the amplitude of each element in the post-state by multiplying a factor $\frac{1}{\sqrt{p}}$, with $r=\Msum_{j=0}^m \slen{z_j}^2$ as the sum of the amplitude squares appearing in the post-state. 
The rule's transition is labeled with $d.p$, referring to the measurement result bitstring $d$ and the probability of having the result.
Measurement operations cause locus scope changes in the quantum state, and \disq ensures the program correctness by our type system in \Cref{sec:typing}.

\begin{figure*}[t]
{\tiny
  \begin{mathpar}       
   \inferrule[S-Comp]{(\Phi, P) \xrightarrow{\xi.p} (\Phi', P')}
       {(\Phi, \sdot{P}{Q}) \xrightarrow{\xi.p} (\Phi', \sdot{P'}{Q})}
       
       \inferrule[S-Move]{n=\slen{R,\overline{T}}\\ (\textcolor{spec}{\{\kappa:\theta\}},R) \xrightarrow{p} (\textcolor{spec}{\{\kappa':\theta'\}}, R')}
       {(\Phi\cupdot \textcolor{spec}{\{\glocus{\kappa}{l}\sqcupplus K : \theta\}}, \scell{R,\overline{T}}_l)
        \xrightarrow{l.\frac{p}{n}} (\Phi\cupdot \textcolor{spec}{\{\glocus{\kappa'}{l}\sqcupplus K: \theta'\}}, \scell{R',\overline{T}}_l)}
        
  \inferrule[S-Mem]{n=\slen{R,\overline{T}}}
       {(\Phi, \scell{\seq{D}{R},\overline{T}}_l)
        \xrightarrow{l.\frac{1}{n}} (\Phi, \sacell{\seq{D}{R}}{\overline{T}}_l)}
        
     \inferrule[S-CommA]{}
       {(\Phi, \sacell{\csenda{a}{v}.R}{{\overline{M}}}_l,\sacell{\creva{a}{x}.T}{\overline{N}}_r)
        \xrightarrow{l.r.1} (\Phi,\scell{{R,\overline{M}}}_l,\scell{{T[v/x]},\overline{N}}_r)}
       
   \inferrule[S-CommC]{\Phi_1 = \textcolor{spec}{\{\glocus{c[0,n)}{l}\sqcupplus \glocus{c[0,n)}{r} : \bigotimes_{k=0}^{n\sminus 1}\sum_{d=0}^1{\frac{1}{\sqrt{2}}\ket{{d}}\ket{d}} \}} }
       {(\Phi\cupdot \Phi_1 \cupdot \textcolor{spec}{\{\glocus{\kappa}{l}\sqcupplus K : \theta\}}, \sacell{\csenda{c(n)}{\kappa}.R}{{\overline{M}}}_l,\sacell{\creva{c(n)}{x}.T}{\overline{N}}_r)
        \xrightarrow{l.r.1} (\Phi \cupdot \textcolor{spec}{\{\glocus{\kappa}{r}\sqcupplus K : \theta\}},\scell{{R,\overline{M}}}_l,\scell{{T[\kappa/x]},\overline{N}}_r)}
    \end{mathpar}
}
\vspace{-1em}
\caption{Selected membrane-level semantic rules.}
\label{fig:exp-semantics-2}
\vspace{-1em}
\end{figure*}

\noindent\textbf{\textit{Membrane Level Semantics.}}
\Cref{fig:exp-semantics-2} shows the selected membrane level semantics. A \disq program is a set of membranes, and every subset of the set can make a move.
The transitions of the processes in a membrane can be understood as a Markov chain, in the sense that every process in a membrane has the chance to be selected to perform a location action or a communication action that requires an airlock step. This indicates that the chance of selecting any of the processes in a membrane equals $\frac{1}{n}$, where $n$ is the number of processes in the membrane. We model the connection between the process and membrane level semantics, via rules \rulelab{S-Mem} and \rulelab{S-Move}.
The former handles the airlock mechanism for selecting a process in a membrane, ready for communication with another membrane, and the latter connects local action transitions with transition behaviors at the membrane level.

In \rulelab{S-Move}, the locus $\glocus{\kappa}{l}\sqcupplus K$ is assumed to map to the data $\theta$ in the quantum state, and the prefixed action in $R$ coincidentally is applied to the quantum datum pointed to by locus $\kappa$ (in membrane $l$), which is guaranteed by the \disq type system.
As we mentioned in \Cref{sec:state}, the qubits mentioned in $\kappa$ might be less than the qubits in the datum valuation $\theta$, as the locus $\kappa$ indicates that only the $\slen{\kappa}$ length qubits prefixed in $\theta$ are able to be manipulated in the process $R$, while the remaining in $\theta$ is unreachable.
The label $l.\frac{p}{n}$ indicates a nondeterministic choice of location $l$, where $p$ is the probability of a one-step move in $R$.
Rule \rulelab{S-Rev} permits the release of an airlock.

Rule \rulelab{S-CommA} performs a classical message communication, from traditional $\pi$-calculus \cite{MILNER19921},
while \rulelab{S-CommC} performs a quantum message communication, assuming that the channel $c\srange{0}{n}$ is properly initialized as an $n$ bitwidth Bell pair, as $\Phi_1$. 
Rules \rulelab{S-CommA}, and \rulelab{S-CommC} transitions have labels $l.r.1$, meaning that the nondeterministic event happens across the $l$ and $r$ membranes.
The probability $1$ in the above three rules ensures that the transitions always happen.

As mentioned in \Cref{sec:probnature}, \disq intends to capture the probabilistic behavior of quantum computation via transition labels.
Rules \rulelab{S-Mem} and \rulelab{S-Move} unveil that a membrane might contain different independent processes executing a series of events, each single event execution happens in the same probability. \Cref{sec:probnature} also shows the probabilistic nature of single-location parallelism.

\subsection{\disq Type System and Metatheories}\label{sec:typing}

The \disq type system is also two-leveled.
The membrane level judgment is $\Omega;\Sigma\vdash \overline{P} \triangleright \Sigma'$, stating that $\overline{P}$ is well-typed under the kind and type environments $\Omega$ and $\Sigma$, producing the post-environment $\Sigma'$.
The process level typing judgment is $\omega; \sigma\vdash \overline{R} \triangleright \sigma'$, stating that $\overline{R}$ is well-typed under the environments $\omega$ and $\sigma$.
The membrane level kind environment $\Omega$ is a map ($l \to x \to g$) and the process level kind environment $\omega$ is a map ($x \to g$). 
The membrane level type environment $\Sigma$ is a locus set, and the process level type environment $\sigma$ is a set of local loci.
The $\cmode$-kind variables in a kind environment $\omega$ are populated through message receipt and quantum measurement operations, while the $\qmodename$-kind variables are populated through a channel $\downc{c}{n}$ and qubit array $\downd{x(n)}$ creation operation.
Selected type rules are in \Cref{fig:exp-sessiontype}.
For every type rule, well-formed domains \((\Omega \vdash \dom{\Sigma})\) (or \((\omega \vdash \dom{\sigma})\)) are required but hidden from the rules, such that every variable used in all loci of \(\Sigma\) (or $\sigma$) must appear in \(\Omega\).
The type system enforces three properties below.  

\begin{figure}[t]
  {
    {\scriptsize
      \begin{mathpar}
\inferrule[T-OP]{\omega;\sigma\cupdot\{\kappa\sqcupplus\kappa'\}\vdash R \triangleright \sigma'}{\omega;\sigma\cupdot\{\kappa\sqcupplus\kappa'\} \vdash \seq{\ssassign{\kappa}{}{\mu}}{R} \triangleright \sigma' }
        
        \inferrule[T-Mea]{\omega[x\mapsto \cmode];\sigma\cupdot\{\kappa'\}\vdash R \triangleright \sigma'}{\omega;\sigma\cupdot\{\kappa\sqcupplus\kappa'\} \vdash \seq{\smea{x}{\kappa}}{R} \triangleright \sigma' }
        
        \inferrule[T-SendQ]{ \{c[0,n)\} \cupdot \{\kappa\} \subseteq \sigma \\ \slen{\kappa} = n\\\\ \omega ; \sigma \textbackslash ( \{c[0,n)\} \cupdot \{\kappa\}) \vdash R \triangleright \sigma'}
        {\omega;\sigma \vdash \seq{\csenda{c(n)}{\kappa}}{R} \triangleright \sigma'}
        
          \inferrule[T-RevQ]{ \{c[0,n)\} \subseteq \sigma  \\\\  \omega[x \mapsto Q(n)] ; \sigma \cupdot\{x[0,n)\} \textbackslash \{c[0,n)\} \vdash R \triangleright \sigma'}
        {\omega;\sigma \vdash \seq{\creva{c(n)}{x}}{R} \triangleright \sigma'}
                
     \inferrule[T-Mem]{\cn{has\_mea}(\overline{R}) \quad \neg\cn{has\_mea}(\overline{T}) \quad \forall j \in [0,\slen{\overline{R}}) . \,\Omega(l);\sigma_j \vdash \overline{R}_j \triangleright \sigma'_j  \quad \Omega(l);\sigma \vdash \overline{T} \triangleright \sigma' \quad \Omega;\Sigma \vdash \overline{Q} \triangleright \Sigma' }
        {\Omega;\glocus{({\biguplus}_{j \in [0,\slen{\overline{R}})} \sigma_j)\cupdot \sigma}{l}
          \cupdot \Sigma \vdash \scell{\overline{R},\overline{T}}_l,\overline{Q} \triangleright \glocus{({\biguplus}_{j \in [0,\slen{\overline{R}})} \sigma'_j)\cupdot \sigma'}{l} \cupdot \Sigma'}
      \end{mathpar}
    }
  }
    \vspace*{-1em}
  {\footnotesize
  \begin{center}
  $\begin{array}{lcl}
  \glocus{\sigma}{l} &\triangleq& \forall \glocus{\kappa}{r}:\tau \in \glocus{\sigma}{l} \,.\, r = l
  \end{array}$
  \end{center}
  }
  \vspace*{-1em}
  \caption{Selected type rules. $\cn{has\_mea}(\overline{R})$ : each $R\in\overline{R}$ has a measurement op.}
  \label{fig:exp-sessiontype}
\vspace*{-1em}
\end{figure}

\noindent\textbf{\textit{Ensuring Proper Parameter Kinds and Scopes.}}
The type system ensures scoping properties for variables and channels; e.g., quantum channels and variables have kind $\qmode{n}$, while classical channels and variables have kind $\cmode$.
Quantum channels and variables must be created before use and can be modified within a membrane. However, they cannot be referenced by operations from distinct membranes. Additionally, some operations, such as message sending and receiving, can only refer to classical variables and channels. All these scoping properties are enforced by the type system. 
Details are in \Cref{sec:semanticsa}.

\noindent\textbf{\textit{Ensuring Proper Locus Partitioning and Locality for the Well-formed Constraints.}}
The constraints in \Cref{def:chanconstrant} are guaranteed by typing for locus structure in \disq.
Our type system also ensures that loci are properly and disjointly partitioned in different membranes, and each membrane refers only to the permitted local loci, e.g. 
rules \rulelab{T-OP} and \rulelab{T-Mea} utilize locus structure, $\textcolor{spec}{\{\kappa\sqcupplus\kappa'\}}$, to guarantee qubit disjointness in the quantum operation.
Rule \rulelab{T-Mem} partitions loci by partitioning type environment, into pieces for different membranes,
and ensures a properly separated analysis of different loci and quantum parameters in different membranes,
where the structure $\textcolor{spec}{\glocus{\sigma}{l}}$ is a subset of the type environment and represents a procedure of collecting all the loci referred to membrane $P$ residing in location $l$, and type check $P$ with the subset $\textcolor{spec}{\glocus{\sigma}{l}}$.

The constraint (2) in \Cref{def:chanconstrant} is guaranteed by rules \rulelab{T-SendQ} and \rulelab{T-RevQ}, which remove quantum information being sent out, e.g., $\textcolor{spec}{\{c[0,n)\} \cupdot \{\kappa\}}$ in \rulelab{T-SendQ}, in the type environment ($\textcolor{spec}{\sigma \textbackslash(\{c[0,n)\} \cupdot \{\kappa\}}$) in the subsequence computation.
Additionally, we also guarantee the bitwidth of the quantum channel ($\textcolor{spec}{c[0,n)}$) is the same as the $\textcolor{spec}{\slen{\kappa}}$. 
In \rulelab{T-RevQ}, when we receive quantum messages, we rename them as $\textcolor{spec}{\{x[0,n)\}}$ and keep them in the type environment.
The guarantees for other constraints are discussed in \Cref{sec:semanticsa}.

The type system also ensures that the measurement locus scopes are properly partitioned and preserved.
In rule \rulelab{T-Mem}, depending on if a process contains measurement operations ($\cn{has\_mea}$), the quantum qubit resource sharing scheme is different.
We collect all the local loci $\textcolor{spec}{({\biguplus}_{j \in [0,\slen{\overline{R}})} \sigma_j)\cupdot \sigma}$ in $l$,
and partition it further into different disjoint union sets. 
For the processes ($\overline{R}$) with measurement operations, we type-check each $R$ with a disjoint set $\sigma_j$.
This forbids $R$ from sharing qubits with other processes.
If a process contains a measurement, it is not suitable for concurrent single-location behavior with other processes, as this would allow different processes to refer to the same measured qubit. 
We permit a shared qubit set $\sigma$ for the processes $\overline{T}$ without measurements.

\noindent\textbf{\textit{Guiding Locus Equivalence and Rewriting.}}\label{sec:locus-eq}
Sometimes, we want to shuffle the order of the locus for a quantum datum so that an operation can be correctly applied. 
The \disq type system maintains the simultaneity of loci across type environments and quantum states through type-guided state rewrites formalized as equivalence relations.
We saw at the Shor's algorithm examples in \Cref{def:shors} that we might need to merge two entanglement groups or rearrange the qubit positions in loci.
Such rewrites are formulated as type equivalence relations, which are associated with simultaneous quantum state rewrites;
the details are introduced in \cite{li2024qafny} and \Cref{sec:state-eq}.
Here, we provide a taste of how such rewrites can happen.
A locus represents a possibly entangled qubit group.
In many cases, we need to utilize the locus information in the type environment to guide the equivalence rewrites of states guarded by the locus.
We associate a state $\varphi$, with a type environment $\sigma$ by sharing the same domain, i.e., $\dom{\varphi}={\sigma}$.
Thus, the environment rewrites ($\preceq$) happening in $\sigma$ gear the state rewrites ($\equiv$) in $\varphi$.
One example rewrite is to add a qubit $x[j\splus 1]$ to a local locus $x\srange{0}{j}$, and rewrite it to $\kappa$ ($x[j\sminus 1]\sqcupplus x\srange{0}{j\sminus 1}\sqcupplus x[j]$), which can also cause the state rewrites happen accordingly as (from left to right):

{\scriptsize
\[\hspace*{-0.5em}
\begin{array}{l@{~}c@{~}lclcl}
\textcolor{spec}{
\{x\srange{0}{j}  \}}
&\textcolor{spec}{\cupdot}& 
\textcolor{spec}{\{x[j] \}
}
&\preceq&
\textcolor{spec}{
\{x\srange{0}{j\,\splus\,1} \}
}
&\preceq&
\textcolor{spec}{
\{\kappa \}
}
\\
\textcolor{spec}{
\{x\srange{0}{j} : \ssum{d=0}{1}{\frac{1}{\sqrt{2}}}{\ket{\overline{d}}\ket{1}}}\}
&
\textcolor{spec}{\cupdot}& 
\textcolor{spec}{ \{x[j] : \ket{0}\}
}
&\equiv&
\textcolor{spec}{
\{x\srange{0}{j\,\splus\, 1} : \ssum{d=0}{1}{\frac{1}{\sqrt{2}}}{\ket{\overline{d}}\ket{1}}\ket{0}\}
}
&\equiv&
\textcolor{spec}{
\{
\kappa : \ssum{d=0}{1}{\frac{1}{\sqrt{2}}}{\ket{1}\ket{\overline{d}}}\ket{0}\}
}
\end{array}
\]
}
  
\noindent\textbf{\textit{The \disq Metatheories.}}\label{sec:theorems}
We prove our type system's preservation with respect to the semantics, assuming well-formedness.
The theorems rely on the definitions of well-formed domains ($\Omega \vdash \Sigma$) and well-formed states ($\Omega;\Sigma \vdash \Phi$), shown in Appx. \cref{sec:semanticsa}.
With the type preservation theorem, we can show that \disq programs respect the constraints in \Cref{def:chanconstrant}.
Type preservation ensures the three properties above and states that the \disq semantics can describe all different quantum operations without losing generality because we can always use the equivalence rewrites to rewrite the locus state in ideal forms.

\begin{theorem}[Type Preservation]\label{thm:type-progress-oqasm}\rm 
If $\Omega \vdash \Sigma$, $\Omega;\Sigma \vdash \overline{P} \triangleright \Sigma'$,
  $(\Phi, \overline{P}) \xrightarrow{\alpha} (\Phi', \overline{P'})$, and $\Omega;\Sigma \vdash \Phi$,
  then there exists $\Omega_1$ and $\Sigma_1$, $\Omega_1;\Sigma_1 \vdash \overline{P'} \triangleright \Sigma'$ and $\dom{\Omega_1}\subseteq \dom{\Omega}$ and $\Sigma_1\subseteq \Sigma$.
\end{theorem}

The term $\dom{\Omega_1}\subseteq \dom{\Omega}$ means that $\forall l. \dom{\Omega_1}(l) \subseteq \dom{\Omega}(l)$. 
The terms $\dom{\Omega_1}\subseteq \dom{\Omega}$ and $\Sigma_1\subseteq \Sigma$ ensure that the post-environments $\Omega_1$ and $\Sigma_1$ are consistent with the pre-environments $\Omega$ and $\Sigma$. With type preservation, we show that every \disq evaluation maintains the constraints in \Cref{def:chanconstrant}.

\begin{corollary}[Constraint Satisfication]\label{thm:constraints}\rm 
Every well-typed $\overline{P}$, as $\Omega \vdash \Sigma$ and $\Omega;\Sigma \vdash \overline{P} \triangleright \Sigma'$, satisfy the constraint in \Cref{def:chanconstrant}, and every program evaluation $(\Phi, \overline{P}) \xrightarrow{\alpha} (\Phi', \overline{P'})$, maintain the constraint satisfaction in $\overline{P'}$. 
\end{corollary}
 \section{\disq Observable Simulation}\label{sec:disq-equiv}

We develop a new simulation that goes beyond classical simulation to capture probabilistic properties, as demonstrated in \Cref{sec:equivcheck}, which has execution being deterministic.
In \disq, we model this behavior using Markov Decision Processes (MDP).
Unlike the CHAM model, where all interactions are nondeterministic, \disq's choice of the membrane is nondeterministic, while the interaction preceding the choice is probabilistic, i.e., the choice of the process that evolves in the non-deterministically selected membrane is probabilistic. Hence, the presence of both nondeterminism and probabilities makes \disq an MDP, where each evolution involves a nondeterministic choice followed by a probabilistic move. 

When using the simulation, the \emph{synchronization points} used to establish equivalence between two programs do not admit a uniform solution. In equating a sequential quantum program with its distributed version, the synchronization points correspond to the partitions of the distributed programs, whereas in a hybrid quantum-classical program, they may be placed after quantum measurements.
In \disq, we extend the syntax to include an additional synchronization point operation $\boxed{d}$, where $d$ is a bitstring representing the label actions to equate in two transition configurations, allowing it to be empty ($\emptyset$). We permit users to set up the synchronization points for equating quantum states. A synchronization point in \disq is defined as $\{\boxed{d}, \emptyset\}$, either a $\boxed{d}$ point operation, or at the end of program execution ($\zero$).
For example in \Cref{def:shors}, we care about the quantum state initialization before the execution in the $r$ location, so we set the synchronization point to be before the $T'$ process in membrane $r$, as $\seq{\seq{\creva{c'(1)}{q}}{\boxed{d}}}{T'}$.

{\small
\begin{center}
$
D ::= \boxed{d} \mid ...
\qquad
(\Phi, \sacell{\seq{\boxed{d}}{R}}{{\overline{M}}}_l)
\xrightarrow{d.1}
(\Phi', \scell{{R,\overline{M}}}_l)
$
\end{center}
}

We now formally define the \disq simulation,
where we are interested in universal path properties, e.g., for all computation paths, the probability of a specific measure result is $p$; such properties enable the construction of equivalence between a quantum program and its distributed version.

Our semantics (\Cref{sec:disq}) describe a labeled transition system $(\Phi,\overline{P}) \xrightarrow{\gamma.p} (\Phi',\overline{P'})$, where $\Phi$ and $\Phi'$ are quantum states, $\overline{P}$ and $\overline{P'}$ are \disq programs, and $\gamma.p$ is a label, where $\gamma$ is either $\xi$ or a bitstring $d$.
We view a pair $(\Phi,\overline{P})$ of quantum state and \disq program as a transition configuration, and permit an additional syntax and semantic rule for the synchronization points above. 

The \disq observable simulation is defined over finite sets of configurations, named as $G$ or $H$, each element in the set has the form $(\Phi,\overline{P})^p$, where $(\Phi,\overline{P})$ is a transition configuration and the probability $p$ is the accumulated probability.
Each program evaluation starts with a root configuration set $\{(\Phi,\overline{P})^1\}$, where $\Phi$ and $\overline{P}$ are the initial state and \disq program.

\begin{definition}[\disq Configuration Set]\label{def:config}\rm
Given a finite transition configuration set $G$, its cardinality $\slen{G}$, and elements $(\Phi_j,\overline{P}_j)^{p_j}$ ($\forall j \in [0,\slen{G})$), we define a syntactic sugar $G^p$, where $p = \sum_{j=0}^{\slen{G}\sminus 1} p_j$. The predicate $\cn{same}(\Phi,G^p)$ states that all elements having the same state $\Phi$, i.e., every element is $(\Phi,\overline{P}_j)^{p_j}$, for all $j$.
\end{definition}

A root configuration set $G^1=\{(\Phi,\overline{P})^1\}$ contains solely the initial program with an initial state $\Phi$.
For every configuration $(\Phi,\overline{P})$ in the set, we can evaluate it and insert its results back into the set, generating a new configuration set $G_1^p$. Clearly, the top program-level configuration has $p=1$ due to the stochasticity of the \disq semantics.
We then define the set of transitions related to the set of transition configurations $G^p$ below.

\begin{definition}[\disq Configuration Set Transition]\label{def:otrans}\rm
Given a transition configuration set $G^p$, we define the set transition $G^p \xrightarrow{\gamma} G_1^{t}$ below.

  \begin{itemize}
  \item for all $(\Phi_j,\overline{P}_j)^{p_j}$ in $G^p$, $(\Phi_j,\overline{P}_j) \xrightarrow{\gamma.t_j} (\Phi'_j,\overline{P'}_j)$, $G_1$ contains all config. $(\Phi'_j,\overline{P'}_j)^{p_j * t_j}$, transitioned from $(\Phi_j,\overline{P}_j)$, and $t=\sum_j p_j * t_j$, for all $j$.
  \end{itemize}
\end{definition}

Since the \disq semantics is stochastic, $p$ is equal to $t$ above, shown below.

\begin{lemma}[Trans. Stochasticity]\label{def:otranslemma}\rm
Given a set $G^p$, if $G^p \xrightarrow{\gamma} G_1^{t}$, then $p=t$.
\end{lemma}

We can now define the \disq observable simulation below.

\begin{definition}[\disq Observable Simulation]\label{def:osim}\rm
Given two configuration sets $G$ and $H$ (written as $G^1$ and $H^1$), $G$ simulates $H$,  written as $G \sqsubseteq H$, iff

  \begin{itemize}
  \item $G=G_1^p \cup G_2^{p'}$, $G_1^p \xrightarrow{d} G_3^{p}$ and $\cn{same}(\Phi,G_1^p)$, if there is $H_1$, $H_2$, and $H_3$, such that $H=H_1^{t} \cup H_2^{t'}$, $H_1^t \xrightarrow{d} H_3^{t}$, $\cn{same}(\Phi,H_1^t)$, $p \approx t$, and $G_3^p \cup G_2^{p'} \sqsubseteq H_3^t \cup H_2^{t'}$.
    
  \item $G=G_1^p \cup G_2^{p'}$ and $G_1^p \xrightarrow{\xi} G_3^{p}$, then $G_3^p\cup G_2^{p'} \sqsubseteq H$.
  
  \item $H=H_1^t \cup H_2^{t'}$ and $H_1^t \xrightarrow{\xi} H_3^{t'}$, then $G \sqsubseteq H_3^t\cup H_3^{t'}$
  \end{itemize}
\end{definition}

One can develop an (on-the-fly) algorithm for observable simulation as a least fixed point computation of the negation of the simulation relation~\cite{BMRRV:ICLP01}.
Instead of computing $G \sqsubseteq H$, we compute $\cn{not\_sim}(\{G\},\{H\})\triangleq \neg (G \sqsubseteq H)$.
Here, we start with two configuration sets, each containing only the initial configurations, i.e., $\overline{G}$ and $\overline{H}$ are respectively initialized as $\{G\}$ and $\{H\}$, as they contain all the possible initial states for the two programs being simulated. In each iteration, we partition a configuration set in the different sets, if the transition configuration set leads to different labels, e.g., in the first iteration, we partition $\overline{G}$ into different sets, such as $G=G_1\uplus G_2 \uplus ...$, for each $G_j$, we guarantee that $G_j^p \xrightarrow{\gamma_j} G_{j1}^{p}$ for one observable label $\gamma_j$.
Then, we check if there is also a partition in $H$, such that $H=H_1\uplus H_2 \uplus ...$, for each $H_j$, we make sure that $H_j^t \xrightarrow{\gamma'_j}H_{j1}^{t}$ for the same label $\gamma'_j$.
For $G_j$, if we cannot find $H_j$, such that $G_j \sqsubseteq H_j$, the $\cn{not\_sim}$ predicate holds.
Otherwise, we loop to check $\cn{not\_sim}(\{G_j\},\{H_j\})$.
We take the least fixed point of the computation, and the negation of its result determines the simulation relation between $G$ and $H$.

We implement a \disq interpreter in Java and the $\cn{not\_sim}$ function on top of our \disq interpreter as our simulation checker.
We then utilize the checker to validate the simulation relation between sequential quantum programs $P$ and their distributed versions $\overline{P'}$, i.e., $\overline{P'} \sqsubseteq P$.
Since $P$ is typically a sequential program, a simulation check is enough to equate the two. Certainly, one can easily construct a bisimulation checker based on our simulation framework for other utilities.
We enable the simulation checks for all case studies in the paper.

 \section{Case Studies}
\label{sec:arith-oqasm}

To demonstrate the \disq utility, we present several case studies, with Shor's algorithm here and more in \Cref{sec:morecases}.
 A key recursive process ($Rec$) used here is defined as $Rec(j,n,f) \triangleq \sifb{j = n}{\zero}{\sact{f(j)}{Rec(j\splus 1)}}$.

\subsection{Distributed Shor's Algorithm}\label{sec:dis-shorsa}

We show the full story of the distributed Shor's algorithm in \Cref{fig:shorseqa,fig:shordisa}.
With the one-step distributed Shor's algorithm definition in \Cref{def:shors}, we show the complete distributed version below.

\begin{example}[Distributed Shor's Algorithm]\label{def:example8}\rm Initially, the membranes $l$ and $u$ respectively holds $n$-qubit $x$ and $y$ qubit arrays, and $r$ hold no qubits.
$x[0,n)$ and $y[0,n)$ have initial states $\textcolor{spec}{\ket{0}}$, respectively.
Membranes $l$ and $u$ share an $n$-qubit quantum channel $c$, while $u$ and $r$ share an $n$-qubit quantum channel $c'$.
\end{example}

{\small
\begin{center}
$
\begin{array}{l}
\textcolor{blue}{\text{Processes: }}
\\
\begin{array}{l@{\;}c@{\;}l@{\qquad\qquad}l@{\;}c@{\;}l}
He(j) &=& \sact{\ssassign{x[j]}{}{\cn{H}}}{\sact{\downd{c(1)}}{\sact{\csenda{c(1)}{x[i]}}{\zero}}} \\
HR(n) &=& Rec(0,n,He)\\[0.8em]
Me(j) &=& \sact{\sact{\downd{c(1)}}{\creva{c(1)}{w}}}{\sact{\ssassign{w\sqcupplus y[0,n)}{}{\cn{CU}(v^{2^{j}}, N)}}{\sact{\downd{c'(1)}}{\sact{\csenda{c‘(1)}{w}}{\zero}}}} \\
MR(n) &=& Rec(0,n,Me)\\[0.8em]
Ed(j) &=& \sact{\downd{c'(1)}}{\sact{\creva{c’(1)}{q[j]}}{\zero}} \\
ER(n) &=& {Rec(0,n,Ed)} \\[0.8em]
\end{array}
\\
\textcolor{blue}{\text{Membranes: }}
\\
\qquad\qquad
{\scell{HR(n)}_l},\,
{\scell{{{MR(n)}}}_u},\,
{\scell{\sact{ER(n)}{{\sact{\ssassign{q[0,n)}{}{\cn{QFT}^{-1}}}{\sact{\smea{d}{q[0,n)}}{ps(d)}}}}}_r}
\end{array}
$
\end{center}
}

The purpose of the distribution is to put $x$ and $y$ qubit arrays in two different machines, so the entangled qubit numbers are limited to $n+1$ in each machine.
To do so, membrane $l$ is responsible to prepare superposition qubits in $x$ array through the $HR$ process;
we apply a \cn{H} gate to $\textcolor{spec}{\glocus{x[j]}{l}}$ and \cn{CX} gate to the $\textcolor{spec}{\glocus{x[j]}{l}}$ and $\textcolor{spec}{\glocus{c[j]}{l}}$ qubits.
Membrane $u$ entangles $x$ and $y$ arrays by executing a loop program through the $MR$ process, i.e., each loop step applies a controlled-$U$ gates between $\textcolor{spec}{\glocus{x[j]}{u}}$ and the $y$ array to entangle these two and then send $\textcolor{spec}{\glocus{x[j]}{u}}$ to $r$.
Membrane $r$ applies the phase estimation step, where it waits for all the qubits from the $x$ array to arrive from $u$, via the $c'$ channel, and then applies $\cn{QFT}^{-1}$ and measurement.

We now explain the communications among the three membranes.
Assuming that $n$ pairs of quantum channels $c(1)$ and $c'(1)$ are created, i.e., one pair of $c(1)$ and $c'(1)$ created at each loop step, the communications among the three membranes are managed by $c$ and $c'$, indicated by the channel edges in \Cref{fig:shordisa}, and they are managed in an $n$-step loop structure.
In each $j$-th loop step, we use one qubit Bell pair in a new quantum channel $c(1)$, connecting $l$ and $u$ as $\textcolor{spec}{\glocus{c[0]}{l}}$ and $\textcolor{spec}{\glocus{c[0]}{u}}$, to transform the information in $x[j]$ in membrane $l$ to $\textcolor{spec}{\glocus{c[0]}{u}}$ in membrane $u$; such a procedure is finished by single qubit teleportation.
The $j$-th loop step also contains several operations in membrane $u$,
Here, we first apply the controlled-$U$ and \cn{CX} gates mentioned above,
and then perform a single qubit teleportation to transform the information in $\textcolor{spec}{\glocus{x[j]}{u}}$ to membrane $r$ via the channel $c'(1)$.
The arrows in \Cref{fig:shordisa} indicate the message passing order of each loop step, including a single qubit teleportation for transforming $\textcolor{spec}{\glocus{x[j]}{l}}$ to $\textcolor{spec}{\glocus{x[j]}{u}}$
and another teleportation for transforming $\textcolor{spec}{\glocus{x[j]}{u}}$ to $\textcolor{spec}{\glocus{x[j]}{r}}$.
Ultimately, we teleport the information of $x[j]$ to membrane $r$.
After the communication loop is executed, we then apply the $\cn{QFT}^{-1}$ and measurement in membrane $r$ to $\textcolor{spec}{\glocus{x[0,n)}{r}}$ at once.
The application $ps(w)$ in \Cref{def:example8} refers to the post-processing step after the quantum order funding step.

In every loop step, $u$ only holds $n+1$ qubits; once the qubit $\textcolor{spec}{\glocus{c[j]}{u}}$ is destroyed after its information is transferred to $r$.
This discussion omits the fact that the modulo multiplication circuit in membrane $u$ might require many more ancillary qubits,
which can be handled based on future circuit distribution, such as the addition circuit distributions in \Cref{sec:dis-adder}.
To equate Shor's algorithm with the distributed version, we have the following proposition.
It is trivial to see that the distributed version simulates the original Shor's algorithm since each membrane above contains only one process, i.e., there is no concurrency, and non-determinism is synchronized by classical message passing.

\begin{theorem}[Distributed Shor's Algorithm Simulation]\label{thm:shors-sim}\rm 
Let Dis-Shors refer to the distributed Shor's program and Shors refer to the sequential one, with two $n$-length input qubit arrays $x$ and $y$, thus, Dis-Shors $\sqsubseteq$ Shors.
\end{theorem}

We verify \Cref{thm:shors-sim} in Coq and utilize the same $\cn{not\_sim}$ simulation checking procedure in \Cref{sec:dis-adder} to automatically validate the theorem.   \section{Related Work}
\label{sec:related}

Many previous studies inspire the \disq development.

\noindent\textbf{Concurrent Quantum Frameworks.}
Many works study quantum concurrency and its verification.
Ying and Feng \cite{4745631} proposed an algebraic logical system for partitioning a sequential quantum program into concurrently executable components.
Feng {\it et al.} \cite{10.1145/3517145} and Ying {\it et al.} \cite{YING2022164,Ying2018} developed proof systems for concurrent quantum programs based on quantum Hoare logic \cite{qhoare}, and Zhang and Ying \cite{zhang2024atomicity} further incorporated atomicity.
Eisert {\it et al.} \cite{PhysRevA.62.052317} showed the resource estimation of implementing a non-local gate theoretically, without investigating how long-distance entanglement can be established.
Ardeshir-Larijani {\it et al.} \cite{10.1007/978-3-642-54862-8_42} developed equivalence checkers for concurrent quantum programs, while \cite{10.1007/978-3-642-36742-7_33} developed an equivalence checker for quantum networking protocols.
These works primarily reason about concurrent behaviors under the assumption that a distributed/concurrent program is already given.
In contrast, \disq focuses on constructing distributed realizations from sequential specifications and validating such rewrites via refinement, while also treating quantum teleportation as a communication mechanism.

\noindent\textbf{Classical and Quantum Process Algebra.}
Traditional process calculi, including CSP \cite{Hoare:1985:CSP:3921} and $\Pi$-calculus \cite{MILNER19921}, provide mature notions of bisimulation and refinement \cite{FDR2,fdr3,DBLP:conf/concur/Sangiori93}, and the Chemical Abstract Machine \cite{BERRY1992217} inspired \disq.
On the other side, a large body of work develops quantum process calculi for describing quantum communication and security protocols, such as qCCS \cite{10.1145/1507244.1507249,10.1145/2400676.2400680}, CQP \cite{10.1145/1040305.1040318}, and QPAlg \cite{10.1145/977091.977108}.
Compared with the above formalisms, \disq models a distributed quantum processor with explicit locations and quantum communication channels.
Moreover, \disq adopts an MDP-based semantic view, enabling a simulation/refinement relation that leverages classical probabilistic verification techniques, rather than defining simulations directly over density matrices.

  \section{Conclusion}\label{sec:limit}

We presented DisQ, a formal model for distributed quantum processors with explicit locations and communication, enabling systematic rewrites of sequential quantum programs into distributed realizations. 
DisQ’s operational semantics and type system provide an MDP view that supports observational refinement checking, and our prototype checker validates the refinements on representative case studies.

\noindent\textbf{Acknowledgments.} This material is based upon work supported by NSF under Award Number CCF2330974.
This paper is dedicated to the memory of our dear co-author Rance Cleaveland.

\ignore{
\begin{acks}                            This material is based upon work supported by NSF under Award Number 2330974.
This paper is dedicated to the memory of our dear co-author Rance Cleaveland.
\end{acks}
}

\bibliography{reference}

\newpage
\appendix
\section{\disq Semantics and Type System and Kind Checking}\label{sec:semanticsa}

\begin{figure*}[h]
{\scriptsize
  \begin{mathpar}
     \inferrule[S-New]{}
       {(\varphi, \downc{x}{n}.R)
        \xrightarrow{1} (\varphi\cupdot \textcolor{spec}{\bigcupdot_{j=0}^{n\sminus 1} \{x[j] : \ket{0} \}}, R)}
               
      \inferrule[S-IFT]{}{ (\varphi,\sifb{1}{R}{T}) \xrightarrow{1} (\varphi,R)}
     
      \inferrule[S-IFF]{}{ (\varphi,\sifb{0}{R}{T}) \xrightarrow{1} (\varphi,T)}
  \end{mathpar}
}
{\footnotesize
\begin{center}
$
\begin{array}{lclcl}
\denote{\mu}^{n}(\sum_{j}{z_j\ket{d_j}\beta_j})
&\triangleq&
 \Msum_{j}{z_j(\denote{\mu}\ket{d_j})\beta_j} & \texttt{where}& \forall j\,\slen{d_j}=n
 \\[0.2em]
 \qfun{(\Msum_{i}{z_i}{\ket{d_{i}}}{\beta_i}+\theta)}{\kappa,b} &\triangleq& \Msum_{i}{z_i}{\ket{d_{i}}}{\beta_i}
      &\texttt{where}&\forall i.\,\slen{d_{i}}=\slen{\kappa}\wedge \denote{b[d_{i}/\kappa]}=\texttt{true}
\end{array}
$
\end{center}
}
\vspace{-0.8em}
\caption{Remaining \disq single process semantic rules.}
\label{fig:exp-semantics-3}
\end{figure*}

\begin{figure*}[t]
{\tiny
  \begin{mathpar}
         \inferrule[S-Rev]{}
       {(\Phi, \sacell{{R}}{\overline{T}}_l)
        \xrightarrow{l.1} (\Phi, \scell{R,\overline{T}}_l)}       
        
 \inferrule[S-End]{}
       {(\Phi, \scell{\overline{\zero}}_l) \xrightarrow{l.1} {(\Phi, \emptyset)}}

       \inferrule[S-NewChan]{}
       {(\Phi, \sacell{\downc{c}{n}.R}{{\overline{M}}}_l, \sacell{\downc{c}{n}.T}{\overline{N}}_r)
        \xrightarrow{l.r.1} (\Phi\cupdot \textcolor{spec}{\bigcupdot_{i=0}^{n\sminus 1}\{\glocus{c[i]}{l}\sqcupplus \glocus{c[i]}{r} : \sum_{d=0}^1{\frac{1}{\sqrt{2}}\ket{{d}}\ket{d}} \}}, \scell{{R,\overline{M}}}_l,\scell{{T[v/x]},\overline{N}}_r)}
  \end{mathpar}
}
\vspace{-1.2em}
\caption{Remaining membrane-level semantic rules.}
\label{fig:exp-semantics-4}
\vspace{-1em}
\end{figure*}

Rule \rulelab{S-New} creates new $n$ blank qubits for locus $x\srange{0}{n}$.
Rule \rulelab{S-IFT} and \rulelab{S-IFF} describe the semantics of classical conditionals.

The subset move behavior is captured by \rulelab{S-Comp}.
In \disq, every membrane has a fixed amount of processes in its lifetime.
In rules \rulelab{S-Mem} and \rulelab{S-Move}, each probabilistic choice of performing a process has a probability $\frac{1}{n}$ where $n$ is the number of processes in the membrane.
To guarantee the equal distribution of the probabilistic choice of a process, we include rule \rulelab{S-Self} in \Cref{fig:exp-semantics-1},
as a $\zero$ process can make a move to itself. 
In the end, if every process in a membrane turns to $\zero$, rule \rulelab{S-End} permits its termination.

Rule \rulelab{S-NewChan} creates a new quantum channel between the membranes $l$ and $r$, which results in $n$ pairs of Bell pair state connecting $l$ and $r$, each of which is pointed to by the locus $\textcolor{spec}{\glocus{c[i]}{l}\sqcupplus \glocus{c[i]}{r}}$ for $i \in [0,n)$.

\begin{figure}[h]
{\small
  \begin{mathpar}
    \inferrule[ ]{\omega(x)=\cmode }{\omega \vdash x : \omega(x)}

    \inferrule[ ]{ \omega \vdash a_1 : C \\ \omega \vdash a_2 : C }{\omega \vdash_l a_1 \,\splus\, a_2 : C}   

    \inferrule[ ]{ \omega \vdash a_1 : C \\ \omega \vdash a_2 : C }{\omega \vdash_l a_1 \cdot a_2 : C}   
 
    \inferrule[ ]{ \omega \vdash a_1 : C \\ \omega \vdash a_2 : C }{\omega \vdash_l {a_1}{=}{a_2} : C}   

    \inferrule[ ]{ \omega \vdash a_1 : C \\ \Omega \vdash a_2 : C }{\Omega \vdash_l {a_1}{<}{a_2} : C}

    \inferrule[ ]{ \omega \vdash b : C}{\omega \vdash \neg b : C}
  \end{mathpar}
}
 \caption{Arith and Bool Kind Checking}
  \label{fig:exp-well-typeda}
\end{figure}

The kind checking procedure $\omega\vdash - : C$ verifies if $-$ is a $\cmode$ kind term, based on the kind checking in \cite{li2024qafny},
and the rules for arithmetic and Boolean expressions are in \Cref{fig:exp-well-typeda}.
The construct $-$ here refers to arithmetic, Boolean equations, or a statement.

\begin{figure}[t]
  {
    {\scriptsize
      \begin{mathpar}
      \hspace*{-0.3em}        
        \inferrule[T-If]{ \omega \vdash B : C \\  \omega ; \sigma \vdash R \triangleright \sigma' \\  \omega ; \sigma \vdash T \triangleright \sigma'}
        {\omega;\sigma \vdash \sif{B}{R}{T} \triangleright \sigma'}
        
      \inferrule[T-BindQ]{ \omega[c\mapsto Q(n)] ; \sigma \cupdot \{c[0,n)\} \vdash R \triangleright \sigma'}
        {\omega;\sigma \vdash \seq{\downd{c(n)}}{R} \triangleright \sigma'}
        
     \inferrule[T-BindQ1]{ \omega[c\mapsto Q(n)] ; \sigma \cupdot \{x[0,n)\} \vdash R \triangleright \sigma'}
        {\omega;\sigma \vdash \seq{\downd{x(n)}}{R} \triangleright \sigma'}
        
        \inferrule[T-SendC]{ \omega \vdash a : C \\ \omega \vdash v : C \\ \omega ; \sigma  \vdash R \triangleright \sigma'}
        {\omega;\sigma \vdash \seq{\csenda{a}{v}}{R} \triangleright \sigma'}
        
        \inferrule[T-RevC]{ \omega \vdash a : C \\  \omega[x\mapsto C] ; \sigma \vdash R \triangleright \sigma'}
        {\omega;\sigma \vdash \seq{\creva{a}{x}}{R} \triangleright \sigma'}
        
     \inferrule[T-Mem]{\cn{has\_mea}(\overline{R})\\\neg\cn{has\_mea}(\overline{T})\\ \forall j \in [0,\slen{\overline{R}}) . \,\Omega(l);\sigma_j \vdash \overline{R}_j \triangleright \sigma'_j 
                         \\\Omega(l);\sigma \vdash \overline{T} \triangleright \sigma' \\\Omega;\Sigma \vdash \overline{Q} \triangleright \Sigma' }
        {\Omega;\glocus{({\biguplus}_{j \in [0,\slen{\overline{R}})} \sigma_j)\cupdot \sigma}{l}
          \cupdot \Sigma \vdash \scell{\overline{R},\overline{T}}_l,\overline{Q} \triangleright \glocus{({\biguplus}_{j \in [0,\slen{\overline{R}})} \sigma'_j)\cupdot \sigma'}{l} \cupdot \Sigma'}
      \end{mathpar}
    }
  }
  {\footnotesize
  \begin{center}
  $\begin{array}{lcl}
  \glocus{\sigma}{l} &\triangleq& \forall \glocus{\kappa}{r}:\tau \in \glocus{\sigma}{l} \,.\, r = l
  \end{array}$
  \end{center}
  }
  \vspace*{-0.8em}
  \caption{Remaining type rules.}
  \label{fig:exp-sessiontype-1}
\vspace*{-1em}
\end{figure}

The Boolean ($\omega\vdash B: C$) and arithmetic ($\omega\vdash v: C$) expression checks above in rules \rulelab{T-IF}, \rulelab{T-SendC}, and \rulelab{T-RevC}, ensure that these expressions can only produce classical results and that their parameters are classical.

In each membrane, a quantum channel and variable creation operations, in \rulelab{T-BindQ} and \rulelab{T-BindQ1}, create new channels and variables, as we push their local locus structures in the type environment, $\textcolor{spec}{\sigma \cupdot \{c[0,n)\}}$ and $\textcolor{spec}{\sigma \cupdot \{x[0,n)\}}$, and carry them in the further type checking. This guarantees the constraint (1) in \Cref{def:chanconstrant}.
Rule \rulelab{T-If} requires the output type environments to be the same for the two branches ($\sigma'$).
This indicates that if one branch contains a measurement on certain loci, the other branch must contain a similar measurement of these loci.

The correctness of our type system in \Cref{sec:typing} is assumed to have well-formed domains below.

\begin{definition}[Well-formed locus domain]\label{def:well-formed-ses}\rm 
  The domain of a environment $\Sigma$ (or state $\Phi$) is \emph{well-formed}, written as
  $\Omega \vdash {\Sigma}$ (or $\dom{\Phi}$), iff for every locus $K\in {\Sigma}$ (or $\dom{\Phi}$):
\begin{itemize}
\item $K$ is disjoint unioned, for every two ranges $\glocus{x\srange{i}{j}}{l}$ and $\glocus{y\srange{i'}{j'}}{l}$ in $K$, $x\srange{i}{j}\cap y\srange{i'}{j'}=\emptyset$.

\item For every range $\glocus{x\srange{i}{j}}{l}\in K$, $\Omega(l)(x)=\qmode{n}$ and $[i,j) \subseteq [0,n)$.
\end{itemize}
\end{definition}

Besides well-formed domain definition, we also require that states (\(\Phi\)) being well-formed ($\Omega;\Sigma\vdash \Phi$), defined as follows. Here, we use $\Sigma(K)$ and $\Phi(K)$ to find the corresponding state entry pointed to by a locus $K'$, such that there exists $K_1\,.\,K'=K \sqcupplus K_1$.

\begin{definition}[Well-formed \disq state]\label{def:well-formed}\rm 
  A state $\Phi$ is \emph{well-formed}, written as
  $\Omega;\Sigma \vdash \Phi$, iff ${\Sigma} = \dom{\Phi}$,
  $\Omega\vdash{\Sigma}$ (all variables in $\Phi$ are in $\Omega$), and:
  \begin{itemize}

  \item For every $K \in {\Sigma}$,
    $\Phi(K)=\sch{m}{z_j}{\ket{c_j}}{\beta_j}$, and for all $j$, $\slen{K}= \slen{c_j}$ and
    $\sum_{j=0}^m\slen{z}^2 = 1$.
  \end{itemize}
\end{definition}

\section{\disq Equivalence Relations}\label{sec:state-eq}

\begin{figure}[h]
\captionsetup[subfigure]{justification=centering}
{\tiny
{\hspace*{-2.3em}
\begin{minipage}[t]{0.49\textwidth}
\subcaption{Environment Equivalence}
\begin{center}
 \[\hspace*{-0.5em}
  \begin{array}{r@{~}c@{~}l}
    \sigma & \preceq & \sigma \\[0.2em]
  \{\emptyset:\tau\} \uplus \sigma &\preceq& \sigma\\[0.2em]
   \{\kappa:\tau\} \uplus \sigma &\preceq& \{\kappa:\tau'\} \uplus \sigma\\
&&\texttt{where}\;\;\tau\sqsubseteq\tau' \\[0.2em]
  \{\kappa_1\sqcupplus s_1 \sqcupplus s_2 \sqcupplus \kappa_2 :\tau\} \uplus \sigma &\preceq& \{\kappa_1\sqcupplus s_2 \sqcupplus s_1 \sqcupplus \kappa_2 : \tau\} \uplus \sigma\\\\[0.2em]
  \{\kappa_1 :\tau\} \uplus \{\kappa_2 :\tau\} \uplus \sigma &\preceq& \{\kappa_1 \sqcupplus \kappa_2 :\tau\} \uplus \sigma \\\\[0.2em]
  \{\kappa_1 \sqcupplus \kappa_2 :\tau\} \uplus \sigma &\preceq& \{\kappa_1 :\tau\} \uplus \{\kappa_2 :\tau\} \uplus \sigma
\\
&&
    \end{array}
  \]
\end{center}
  \label{fig:env-equiv}
\end{minipage}
\vline height -8ex
\begin{minipage}[t]{0.49\textwidth}
\subcaption{State Equivalence}
\begin{center}
   \[\hspace*{-0.5em}
   \begin{array}{r@{~}c@{~}l}
    \varphi & \equiv & \varphi \\[0.2em]
  \{\emptyset:q\} \uplus \varphi &\equiv& \varphi\\[0.2em]
   \{\kappa:q\} \uplus \varphi &\equiv& \{\kappa:q'\} \uplus \varphi\\
   &&\texttt{where}\;\;q\equiv_{\slen{\kappa}}q' 
   \\[0.2em]
  \{\kappa_1\sqcupplus s_1 \sqcupplus s_2 \sqcupplus \kappa_2 :q\} \uplus \varphi &\equiv& \{\kappa_1\sqcupplus s_2 \sqcupplus s_1 \sqcupplus \kappa_2 : q'\} \uplus \varphi
\\
&&\texttt{where}\;\;q'=q^{\slen{\kappa_1}}\langle \slen{s_1}\asymp \slen{s_2} \rangle
\\[0.2em]
  \{\kappa_1 :q_1\} \uplus \{\kappa_2 :q_2\} \uplus \varphi &\equiv& \{\kappa_1 \sqcupplus \kappa_2 :q'\} \uplus \varphi 
\\
&&\texttt{where}\;\;q'= q_1\bowtie q_2
   \\[0.2em]
  \{\kappa_1 \sqcupplus \kappa_2 :\varphi\} \uplus \sigma &\equiv& \{\kappa_1 :\varphi_1\} \uplus \{\kappa_2 :\varphi_2\} \uplus \sigma
\\
&&\texttt{where}\;\;\varphi_1 \bowtie \varphi_2=\varphi\wedge \slen{\varphi_1}=\slen{\kappa_1}
    \end{array}
 \]
\end{center}
  \label{fig:qafny-stateequiv}
\end{minipage}
{\tiny
\[\hspace*{-1em}
\begin{array}{l}
\textcolor{blue}{\text{Permutation:}}\\[0.5em]
{
\hspace*{-0.2em}
\begin{array}{r@{~}c@{~}l@{~}c@{~}l}
(q_1 \bigotimes q_2 \bigotimes q_3 \bigotimes q_4)^{n}\langle i\asymp k\rangle
&\triangleq&
q_1 \bigotimes q_3 \bigotimes q_2 \bigotimes q_4
&\;\;\;\texttt{where}&\;\;\slen{q_1}=n\wedge \slen{q_2}=i\wedge \slen{q_3}=k\\[0.5em]
(\Msum_{j}{z_j\ket{c_j}\ket{c'_j}\ket{c''_j}\eta_j})^{n}\langle i\asymp k\rangle
&\triangleq&
\Msum_{j}{z_j\ket{c_j}\ket{c''_j}\ket{c'_j}\eta_j}
&\;\;\;\texttt{where}&\;\;\slen{c_j}=n\wedge \slen{c'_j}=i\wedge \slen{c''_j}=k
\end{array}
}
\\[1.4em]
\textcolor{blue}{\text{Join Product:}}\\
{
\setlength\arraycolsep{4pt}
\begin{array}{r@{~}c@{~}l@{~}c@{~}l r@{~}c@{~}l@{~}c@{~}l}
z_1\ket{c_1}&\bowtie& z_2\ket{c_2} &\triangleq& (z_1 \cdot z_2)\ket{c_1}\ket{c_2}
&
\sch{n}{z_j}{\ket{c_j}}&\bowtie&\schk{m}{z_k}{c_k} &\triangleq& \schi{n\cdot m}{z_j\cdot z_k}{c_j}\ket{c_k}
\\[0.5em]
\ket{c_1}&\bowtie&\Msum_{j}{z_j \eta_j}&\triangleq&\Msum_{j}{z_j\ket{c_1}\eta_j}
&
(\ket{0}+\alpha(r)\ket{1})&\bowtie&\Msum_{j}{z_j \eta_j}
&\triangleq&
\Msum_{j}{z_j\ket{0}\eta_j}+\Msum_{j}{(\alpha(r)\cdot z_j)\ket{1}\eta_j}
\end{array}
}
\end{array}
\]
}
  \caption{\disq type/state relations. $\cdot$ is math mult. Term $\Msum^{n\cdot m}P$ is a summation omitting the indexing details. $\bigotimes$ expands a $\thadt$ array, as $\frac{1}{\sqrt{2^{n+m}}}\Motimes_{j=0}^{n\splus m\sminus 2}q_j=(\frac{1}{\sqrt{2^{n}}}\Motimes_{j=0}^{n\sminus 1}q_j)\bigotimes(\frac{1}{\sqrt{2^{m}}}\Motimes_{j=0}^{m\sminus 1}q_j)$.}
  \label{fig:qafny-eq}
}
}
\end{figure}

\begin{figure}
\captionsetup[subfigure]{justification=centering}
{\scriptsize
{
\begin{minipage}[b]{0.35\textwidth}
\begin{center}
 \[
  \begin{array}{l@{~}cl}
  \tau & \sqsubseteq & \tau \\[0.3em]
  \tnort &\sqsubseteq& \tcht\\[1.3em]
  \thadt &\sqsubseteq& \tcht
    \end{array}
  \]
\end{center}
\subcaption{Subtyping}
  \label{fig:qafny-subtype}
\end{minipage}
\begin{minipage}[b]{0.5\textwidth}
\begin{center}
   \[\hspace*{-3em}
   \begin{array}{l@{~}cl}
   q & \equiv_{\slen{q}} & q \\[0.3em]
  \ket{c} &\equiv_n& \sch{0}{ }{\ket{c}}\\[0.5em]
  \shad{2^n}{n\sminus 1}{\alpha(r_j)} &\equiv_n& \sch{2^n\sminus 1}{\frac{\alpha(\Msum_{k=0}^{n\sminus 1} r_k \cdot \tob{j}[k])}{\sqrt{2^n}}}{\ket{j}}
    \end{array}
 \]
\end{center}
\subcaption{Quantum Value Equivalence}
  \label{fig:qafny-sequiv}
\end{minipage}
\hfill{}
\begin{minipage}[t]{\textwidth}
{\tiny
\[\hspace{-1em}
\begin{array}{r @{~} c @{~} l @{\quad}  r@{~}c @{~} l @{\quad}  r @{~}c @{~} l @{\quad} r @ {~} c @{~} l @{\quad} r @{~} c @{~} l}
x[n,n) & \equiv & \emptyset
&
\emptyset \sqcupplus \kappa & \equiv & \kappa
&
\ket{d_1}\ket{d_2} &\equiv& \ket{d_1 d_2}
&
\glocus{q\sqcupplus q'}{l} &\equiv& \glocus{q}{l}\sqcupplus\glocus{q'}{l}
&
x[n,m) & \equiv & x[n,j)\sqcupplus x[j,m) \,\;\cn{if}\;j \in [n,m]
    \end{array}
  \]
  }
\subcaption{locus Equivalence}
  \label{fig:qafny-ses-equal}
\end{minipage}

  \caption{\disq type/state relations.}
  \label{fig:qafny-eq1}
}
}
\end{figure}

\begin{figure}[h]
{
  \begin{mathpar}
       \inferrule[T-Par]{\sigma \preceq \sigma' \;\quad \Omega;\sigma' \vdash e \triangleright \sigma''}{\Omega;\sigma_1\cupdot\sigma \vdash e \triangleright \sigma_1\cupdot\sigma'' }

        \inferrule[T-ParM]{\Sigma \preceq \Sigma' \;\quad \Omega;\Sigma' \vdash P \triangleright \Sigma''}{\Omega;\Sigma_1\cupdot\Sigma \vdash P \triangleright \Sigma_1\cupdot\Sigma'' }
      \end{mathpar}
}
  \caption{Additional \disq Type Rules.}
  \label{fig:vqimpa}
\end{figure}

The \disq type system maintains simultaneity through the type-guided state rewrites, formalized as equivalence relations (\Cref{fig:qafny-eq1}).
The equivalence relations happen in our type rules \rulelab{T-Par} and \rulelab{T-ParM} in \Cref{fig:vqimpa}.
We only show the rewrite rules for local loci, and the loci with membrane structures can be manipulated through the merged rules in \Cref{fig:disq-state}, as well as a similar style of permutation rules in \Cref{sec:locus-eq}.
Other than the locus qubit position permutation being introduced, the types below associated with loci in the environment also play an essential role in the rewrites.

{\small
\[
\begin{array}{llcl@{\quad}c@{\quad}l@{\quad}c@{\quad}l} 
      \text{Quantum Type} & \tau & ::= & \tnort &\mid& \thadt  &\mid& \tcht \\
      \text{Quantum Value (Forms)} & q & ::= & w & \mid& \shad{2^n}{n\sminus 1}{\alpha(r_j)} &\mid& \ssum{j=0}{m}{w_j}
\end{array}
\]
}

The \disq type system is inherited from the \qafny type system \cite{li2024qafny} with three different types.
Quantum values are categorized into three different types: \(\tnort\), \(\thadt\) and \(\tcht\).
A \emph{normal} value (\(\tnort\)) is an array (tensor product) of single-qubit values \(\ket{0}\) or \(\ket{1}\).
Sometimes, a (\(\tnort\))-typed value is associated with an amplitude $z$, representing an intermediate partial program state.
A \emph{Hadamard} (\(\thadt\)) typed value represents a collection of qubits in superposition but not entangled,
i.e., an $n$-qubit array $\frac{1}{\sqrt{2}}(\ket{0} + \alpha(r_0)\ket{1})\otimes ... \otimes \frac{1}{\sqrt{2}}(\ket{0} + \alpha(r_{n-1})\ket{1})$, can be encoded as $\shad{2^n}{n\sminus 1}{\alpha(r_j)}$, with $\alpha(r_j)=e^{2\pi i r_j}$ ($r_j \in \mathbb{R}$) being the \emph{local phase},
a special amplitude whose norm is $1$, i.e., $\slen{\alpha(r_j)}=1$. 
The most general form of $n$-qubit values is the \emph{entanglement} (\(\tcht\)) typed value,
consisting of a linear combination (represented as an array) of basis-kets, as $\sch{m}{z_j}{\beta_j\eta_j}$, where $m$ is the number of elements in the array.
In \disq, we \textit{extend} traditional basis-ket structures in the Dirac notation to be the above form, so each basis-ket of the above value contains not only an amplitude $z_j$ and a basis $\beta_j$ but also a frozen basis stack $\eta_j$, storing bases not directly involved in the current computation.
Here, $\beta_j$ can always be represented as a single $\ket{c_j}$ by the equation in \Cref{fig:disq-state}.
Every $\beta_j$ in the array has the same cardinality, e.g., if $\slen{c_0}=n$ ($\beta_0=\ket{c_0}$), then $\slen{c_i}=n$ ($\beta_j=\ket{c_j}$) for all $j$.

In \disq, a locus represents a possibly entangled qubit group. From the study of many quantum algorithms \cite{qft-adder,ChildsNAND,mike-and-ike,shors,1366221,Rigolin_2005,10.5555/1070432.1070591,ccx-adder}, we found that the establishment of an entanglement group can be viewed as a loop structure of incrementally adding a qubit to the group at a time, representing the entanglement's scope expansion.
This behavior is similar to splits and joins of array elements if we view quantum states as arrays.
However, joining and splitting two $\tcht$-typed values are hard problems \footnote{The former is a Cartesian product; the latter is $\ge$ $\mathit{NP}$-hard. }.
Another critical observation in studying many quantum algorithms is that the entanglement group establishment usually involves splitting a qubit in a $\tnort$/$\thadt$ typed value and joining it to an existing $\tcht$ typed entanglement group.
We manage these join and split patterns type-guided equations in \disq, suitable for automated verification.

The semantics in \Cref{fig:exp-semantics-1} assumes that the loci in quantum states can be in ideal forms, e.g., rule \rulelab{S-OP} assumes that the target locus $\kappa$ are always prefixed.
This step is valid if we can rewrite (type environment partial order $\preceq$) the locus to the ideal form through rule $\textsc{T-Par}$ and \rulelab{T-ParM} in \Cref{fig:exp-sessiontype}, which interconnectively rewrites the locus appearing in the state, through our state equivalence relation ($\equiv$), as the locus state simultaneity enforcement.
The state equivalence rewrites have two components. 

First, the type and quantum value forms have simultaneity in \Cref{fig:qafny-eq1}, i.e., given a type $\tau_1$ for a locus $\kappa$ in a type environment ($\Sigma$), if it is a subtype ($\sqsubseteq$) of another type $\tau_2$, $K$'s value $q_1$ in a state ($\Phi$) can be rewritten to $q_2$ that has the type $\tau_2$ through state equivalence rewrites ($\equiv_n$) where $n$ is the number of qubits in $q_1$ and $q_2$. Both $\sqsubseteq$ and $\equiv_n$ are reflexive and types $\tnort$ and $\thadt$ are subtypes of $\tcht$, which means that a $\tnort$ typed value ($\ket{c}$) and a $\thadt$ typed value ($\shad{2^n}{n\sminus 1}{\alpha(r_j)}$) can be rewritten to an $\tcht$ typed value. For example, a $\thadt$ typed value $\shad{2^n}{n\sminus 1}{}$ can be rewritten to an $\tcht$ type as $\schii{2^{n}\sminus 1}{\frac{1}{\sqrt{2^{n}}}}{i}$.
If such a rewrite happens, we correspondingly transform $x[0,n)$'s type to $\tcht$ in the type environment. 

Second, type environment partial order ($\preceq$) and state equivalence ($\equiv$) also have simultaneity in \Cref{fig:qafny-eq} for local loci,
and the relations between loci can be derived based on the following rules, as well as permutations on $\uplus$ operations.

{\small
  \begin{mathpar}
    \inferrule[ ]{ \sigma \preceq \sigma' }{\glocus{\sigma \uplus \sigma_1}{l} \uplus \Sigma \preceq \glocus{\sigma' \uplus \sigma_1}{l} \uplus \Sigma}
    
    \inferrule[ ]{ \varphi \preceq \varphi' }{\glocus{\varphi \uplus \varphi_1}{l} \uplus \Phi \preceq \glocus{\varphi' \uplus \varphi_1}{l} \uplus \Phi}
  \end{mathpar}
}

Here, we associate a state $\Phi$, with the type environment $\Sigma$ by sharing the same domain, i.e., $\dom{\Phi}=\dom{\Sigma}$.
Thus, the environment rewrites ($\preceq$) happening in $\Sigma$ gear the state rewrites in $\Phi$. 
In \Cref{fig:qafny-eq}, the rules of environment partial order and state equivalence are one-to-one corresponding. 
The first three lines describe the properties of reflective, identity, and subtyping equivalence.
The fourth line enforces that the environment and state are close under locus permutation.
After the equivalence rewrite, the position bases of ranges $s_1$ and $s_2$ are mutated by applying the function $q^{\slen{\kappa_1}}\langle \slen{s_1}\asymp \slen{s_2} \rangle$.
One example is the following local locus rewrite from left to right, where we permute the two ranges $x\srange{0}{n}$ and $y\srange{0}{n}$.

{\footnotesize
\begin{center}
$\begin{array}{lcl}
\big{\{}x\srange{0}{n}\sqcupplus y\srange{0}{n} : \tcht \big{\}} &\preceq& \big{\{}y\srange{0}{n}\sqcupplus x\srange{0}{n} : \tcht \big{\}} 
\\[0.2em]
\big{\{}x\srange{0}{n}\sqcupplus y\srange{0}{n} : \textcolor{spec}{\schai{2^{n}\sminus 1}{\frac{1}{\sqrt{2^{n}}}}{i}{\ket{a^{i}\;\mmod\;N}}} \big{\}}
&\equiv& 
\big{\{} y\srange{0}{n}\sqcupplus x\srange{0}{n} : \textcolor{spec}{\schai{2^{n}\sminus 1}{\frac{1}{\sqrt{2^{n}}}}{a^{i}\;\mmod\;N}{\ket{i}}}\big{\}}
\end{array}
$
\end{center}
}

The last two lines in \Cref{fig:env-equiv,fig:qafny-stateequiv} describe locus joins and splits, where the latter is an inverse of the former but much harder to perform practically.
In the most general form, joining two $\tcht$-type states computes the Cartesian product of their basis-kets, shown in the bottom of \Cref{fig:qafny-eq};
such operations are computational expensive in verification and validation.
Fortunately, the join operations in most quantum algorithms are between a $\tnort$/$\thadt$ typed and an $\tcht$-typed state, 
Joining a $\tnort$-typed and $\tcht$-typed state puts extra qubits in the
right location in every basis-ket of the $\tcht$-typed state.  

\section{More Case Studies}\label{sec:morecases}

We provide more case studies here.

\subsection{Quantum Teleporation For Ensuring Entanglement Information}\label{sec:controled-ghz}

Quantum teleportation is a quantum network protocol that teleports information about a qubit to remote locations.
This section shows a general use of quantum teleporation to teleport entanglement information.
A key observation is that quantum entanglement is also a piece of information; thus, when teleporting a qubit,
the possible entanglement associated with the qubit should also be kept by remote qubits.

To demonstrate the case, we use the processes in \Cref{def:example0}, as a membrane structure shown in \Cref{def:example5} to teleport a qubit $x[1]$ --- currently entangles with $x[0]$ --- from membrane $l$ to $r$.
The program first creates a shared quantum channel $c(1)$ between the two membranes, referred to by $\textcolor{spec}{\glocus{c[0]}{l}\sqcupplus \glocus{c[0]}{r}}$,
and then teleport $x[1]$ to membrane $r$ to store the information in $\cglocus{x[1]}{r}$.
The result should show that the entanglement between $\cglocus{x[0]}{l}$ and $\cglocus{x[1]}{r}$ is transferred to be an entanglement between $\cglocus{x[0]}{l}$ and $\cglocus{x[1]}{r}$.

\begin{example}[Quantum Teleportation Entanglement Preservation]\label{def:example5}\rm 
The example has two membranes. The program code of membrane $l$ is: ${\scell{\sact{\downc{c}{1}}{\sact{\csenda{c(1)}{x[1]}}{R}}}_l}$, and The program code of membrane $r$ is: ${\scell{\sact{\downc{c}{1}}{\sact{\creva{c(1)}{w}}}{T}}_r}$.    

Membrane $l$ has initially two qubit entangled state $\textcolor{spec}{x[0,2) : z_0\ket{00}+z_1\ket{11}}$.

{\begin{center}
$
\qquad \textcolor{spec}{K_e=\glocus{x[0,2) \sqcupplus c[0]}{l} \sqcupplus \glocus{c[0]}{r}}$
\end{center}
}

The following provides the first few transition steps, where $\neg b$ is the bit-flip of the bit $b$.
The $R$ process in steps (3) and (4) refers to:
\[R  =  {\sact{\smea{i}{b_1}{x[1]}}
       {\sact{\smea{i}{b_2}{c[0]}}
         {\sact{\csenda{a}{b_1}}{\sact{\csenda{a}{2}}{0}}}}}
\].

{\tiny
\begin{center}
$\hspace*{-1em}
\begin{array}{c@{\;\;}c@{\;\;}l}
\textcolor{blue}{
(1)}
&
&
\big{(}
\textcolor{spec}{\{ \glocus{x[0,2)}{l} : \sum_{b=0}^1 z_b\ket{bb}\}},\;
{\scell{\sact{\downc{c}{1}}{\sact{\csenda{c(1)}{x[1]}}{R}}}_l},
{\scell{\sact{\downc{c}{1}}{\sact{\creva{c(1)}{w}}{T}}}_r}
\big{)}
\\[0.4em]
\textcolor{blue}{
(2)}
&
\xrightarrow{l.1}
&
\big{(}
\textcolor{spec}{\{ \glocus{x[0,2)}{l} : \sum_{b=0}^1 z_b\ket{bb}\}},\;
{\sacell{\sact{\downc{c}{1}}{\sact{\csenda{c(1)}{x[1]}}{R}}}{\emptyset}_l},
{\scell{\sact{\downc{c}{1}}{\sact{\creva{c(1)}{w}}{T}}}_r}
\big{)}
\\[0.4em]
\textcolor{blue}{
(3)}
&
\xrightarrow{r.1}
&
\big{(}
\textcolor{spec}{\{ \glocus{x[0,2)}{l} : \sum_{b=0}^1 z_b\ket{bb}\}},\;
{\sacell{\sact{\downc{c}{1}}{\sact{\csenda{c(1)}{x[1]}}{R}}}{\emptyset}_l},
{\sacell{\sact{\downc{c}{1}}{\sact{\creva{c(1)}{w}}{T}}}{\emptyset}_r}
\big{)}
\\[0.4em]
\textcolor{blue}{
(4)}
&
\xrightarrow{l.r.1}
&
\big{(}
\textcolor{spec}{\{ \glocus{x[0,2)}{l} : \sum_{b=0}^1 z_b\ket{bb}, \textcolor{spec}{\glocus{c[0]}{l}\sqcupplus \glocus{c[0]}{r}}  : \frac{1}{\sqrt{2}}\sum_{b=0}^1\ket{bb}\}},
{\scell{\sact{\csenda{c(1)}{x[1]}}{R}}_l},{\scell{\sact{\creva{c(1)}{w}}{T}}_r}
\big{)}
\\[0.4em]
\textcolor{blue}{
(5)}
&
\xrightarrow{l.1}
&
\big{(}
\textcolor{spec}{\{ \glocus{x[0,2)}{l} : \sum_{b=0}^1 z_b\ket{bb}, \textcolor{spec}{\glocus{c[0]}{l}\sqcupplus \glocus{c[0]}{r}}  : \frac{1}{\sqrt{2}}\sum_{b=0}^1\ket{bb}\}},
{\sacell{\sact{\csenda{c(1)}{x[1]}}{R}}{\emptyset}_l},{\scell{\sact{\creva{c(1)}{w}}{T}}_r}
\big{)}
\\[0.4em]
\textcolor{blue}{
(6)}
&
\xrightarrow{r.1}
&
\big{(}
\textcolor{spec}{\{ \glocus{x[0,2)}{l} : \sum_{b=0}^1 z_b\ket{bb}, \textcolor{spec}{\glocus{c[0]}{l}\sqcupplus \glocus{c[0]}{r}}  : \frac{1}{\sqrt{2}}\sum_{b=0}^1\ket{bb}\}},
{\sacell{\sact{\csenda{c(1)}{x[1]}}{R}}{\emptyset}_l},{\sacell{\sact{\creva{c(1)}{w}}{T}}{\emptyset}_r}
\big{)}
\\[0.4em]
\textcolor{blue}{
(7)}
&
\xrightarrow{l.r.1}
&
\big{(}
\textcolor{spec}{\{\textcolor{spec}{\glocus{x[0]}{l}\sqcupplus \glocus{x[1]}{r}} : \sum_{b=0}^1 z_b\ket{bb} \}},
{\scell{R}_l}, {\scell{T[x[1]/w]}_r}
\big{)}
\end{array}
$
\end{center}
}
\end{example}

The above example shows that a Bell pair $\cglocus{x[0,2)}{l}$, which is a quantum channel itself, can be teleported to another membrane, indicated by the locus $\textcolor{spec}{\glocus{x[0]}{l}\sqcupplus \glocus{x[1]}{r}}$.
\disq is able to demonstrate the behavior.
The Bell pair telportation is named quantum entanglement swap, and it is useful in performing long distance quantum message transmission.
The way is to first teleport a half of a Bell pair to a remote location to establish a quantum channel that has longer distance.

\subsection{Parallel Adder}\label{sec:parallel-adder}

\begin{wrapfigure}{r}{3.4cm}
{\scriptsize
  $
  \Qcircuit @C=0.5em @R=0.5em {
    &                                   & &     \qw  & \qw   & \multigate{4}{\texttt{${x[0,i)} + n$}}  & \qw &  \qw \\
    & \push{{x[0,i)}\;\,} & &     \qw  &  \qw & \ghost{\texttt{${x[0,i)} + n$}}          & \qw &   \qw \\
    &                                   & &        &   &                                             & & \\
    &                                   & &       & \dots    &                                       & &  \\
    &                                   & &    \qw  & \qw    &  \ghost{\texttt{${x[0,i)} + n$}}         & \qw &  \qw \\
    &                                   & & \qw &  \qw & \multigate{4}{\texttt{${y[0,j)} - n$}}         & \qw & \qw \\
    & \push{{y[0,j)}\;\,} & & \qw & \qw & \ghost{\texttt{${y[0,j)} - n$}}                & \qw &   \qw \\
    &                                    & &  & \dots  &                                             & & \\
    &                                   & &    &    &                                               & &  \\
    &                                   & & \qw  & \qw  & \ghost{\texttt{${y[0,j)} - n$}}              & \qw & \qw
    \gategroup{1}{3}{5}{3}{1em}{\{}
    \gategroup{6}{3}{10}{3}{1em}{\{}
    }
$  }
\caption{Parallel adders.}\label{fig:paralleladder}
\end{wrapfigure}

\Cref{sec:disq-equiv} describes the \disq simulation, suitable for the analysis of general distributed quantum programs.
In general, quantum programs might contain measurements involving probabilistic behaviors, e.g., each call to the order finding component in \Cref{fig:shors-full} has a probability of success. Single-location quantum computation in a membrane might contain parallelism, emitting a probabilistic choice among different parallelized single-membrane processes. We need a new simulation relation to explore these behaviors.
In implementing the order finding algorithm, the membrane $u$ contains modular-multiplication circuits, having a long circuit depth.
Parallelizing the circuits can improve the performance. We demonstrate a simple example representing the parallelization and performance improvement below.

\begin{example}[Parallel Adder]\label{def:example2}\rm
We define a parallel adder in \Cref{fig:paralleladder}. Two quantum arrays $x[0,i)$ and $y[0,j)$ are entangled in a same membrane, as $\textcolor{spec}{x[0,i)\sqcupplus y[0,j)}$.
We apply a quantum addition ${x[0,i)} + n$ to the range $x[0,i)$ and a subtraction to the range $y[0,j)$. 

The sequential version has one process:
\[\bscell{\seq{\ssassign{x[0,i)}{}{x[0,i) + n}}{\seq{\ssassign{y[0,j)}{}{y[0,j) - n}}{\zero}}}_l\].

The parallel version has two processes: 
\[\bscell{\sdot{\seq{\ssassign{x[0,i)}{}{x[0,i) + n}}{\zero}}{{\seq{\ssassign{y[0,j)}{}{y[0,j) - n}}{\zero}}}}_l\].
\end{example}

Below are two possible parallel program transitions, demonstrating the probabilistic nature of a single membrane due to parallel processes in a membrane $l$.

{\tiny
\[\hspace*{-1.5em}
\begin{array}{c@{\;}c}
\begin{array}{c@{\;\;}l}
&
\bscell{\sdot{\seq{\ssassign{x[0,i)}{}{x[0,i) + n}}{\zero}}{{\seq{\ssassign{y[0,j)}{}{y[0,j) - n}}{\zero}}}}_l
\\
\xrightarrow{l.\frac{1}{2}}
&
\bscell{\sdot{{\zero}}{{\seq{\ssassign{y[0,j)}{}{y[0,j) - n}}{\zero}}}}_l
\\
\xrightarrow{l.\frac{1}{2}}
&
\bscell{\sdot{{\zero}}{{{\zero}}}}_l \xrightarrow{l.1} \emptyset
\end{array}
&
\begin{array}{c@{\;\;}l}
&
\bscell{\sdot{\seq{\ssassign{x[0,i)}{}{x[0,i) + n}}{\zero}}{{\seq{\ssassign{y[0,j)}{}{y[0,j) - n}}{\zero}}}}_l
\\
\xrightarrow{l.\frac{1}{2}}
&
\bscell{\sdot{\seq{\ssassign{x[0,i)}{}{x[0,i) + n}}{\zero}}{{{\zero}}}}_l
\\
\xrightarrow{l.\frac{1}{2}}
&
\bscell{\sdot{\zero}{{\zero}}}_l  \xrightarrow{l.1} \emptyset
\end{array}
\end{array}
\]
}

The above transitions alternatively select the left and right processes in the parallel adder program, resulting in a $\frac{1}{2}$ probability label in each selection.
The quantum operations are single-location and local to the membrane, so no airlock mechanism is involved and the probability label calculation is $\frac{1}{n}$ with $n$ being the number of local processes in the membrane. 

As demonstrated above, \disq utilizes membrane location labels to perform non-deterministic choices of selecting a membrane for transitions, and probabilistic labels to model single-location parallelism, i.e., two quantum operations might apply to completely different and disjoint qubits in a single location, and they can be executed in parallel.

\begin{wrapfigure}{l}{3cm}
\vspace*{-1em}
  \includegraphics[width=0.22\textwidth]{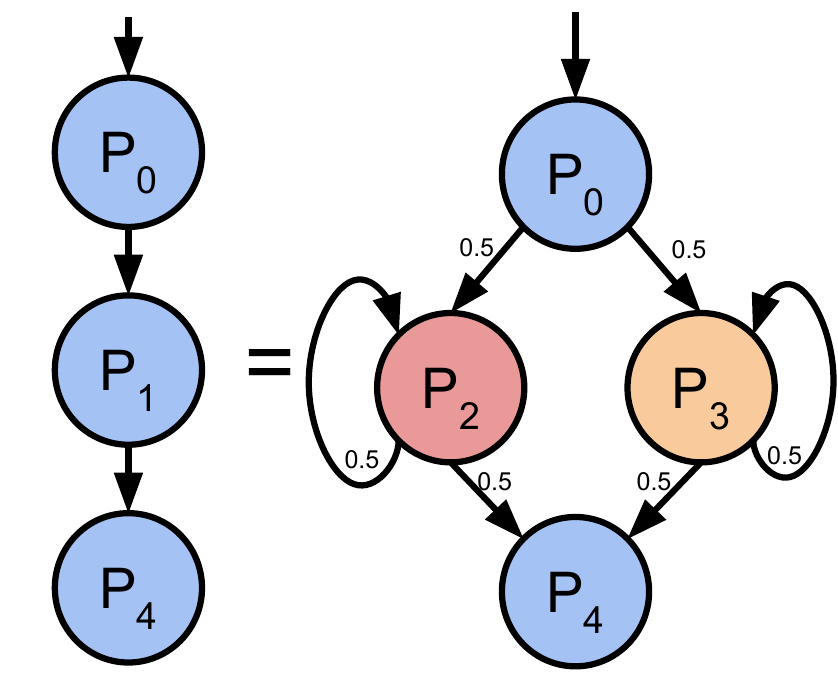}
  \caption{Sim Diagram.}
\label{fig:equiv-sim}
\end{wrapfigure} 

If we sequentialize the above parallel program, the two final states are the same as executing sequential and parallel versions, demonstrated as automata in \Cref{fig:equiv-sim}.
The two transitions above represent a Markov chain, and each transition has a probability $\frac{1}{4}$ of reaching the final state, while the probability along the sequential program execution is $1$. To equate the two programs, we need to sum the probabilities of all different parallel execution paths in the parallel version, e.g., the probabilities in the paths towards $\cn{P}_4$ are summed to $1$ in the two automata in \Cref{fig:equiv-sim}, maintaining stochasticity.
On the other hand, it is clear that equating only the final quantum states of two programs might not be enough for different quantum programs. 
For example, one might want to verify that the step-by-step procedure is equivalent between the two versions of order finding.

We develop the \disq observable simulation to connect the semantics with classical MDP relation, to explore probabilistic behaviors.
Such simulation relation can be conducted based on the sum of different execution paths reaching the same intermediate states that happen at a set of synchronization points.
We provide the ability for users to define synchronization points as the intermediate state locations for performing equivalence checking.
In \Cref{sec:disq}, we define the \disq semantics by viewing distributed quantum systems as a transition system based on MDPs, and we extend the above simulation relation to one in \Cref{sec:disq-equiv}.

\subsection{Distributed QFT Adder}\label{sec:qft-adder}

 \begin{figure}[t]
   \begin{minipage}[b]{0.6\textwidth}
 \[
 \begin{array}{c}
     \tiny
     \Qcircuit @C=0.5em @R=0.75em {
       \lstick{\ket{a_{n-1}}} & \qw & \ctrl{5} & \qw & \qw & \qw & \qw & \qw & \qw & \qw & \rstick{\ket{a_{n-1}}} \\
       \lstick{\ket{a_{n-2}}} & \qw & \qw & \ctrl{4} & \qw & \qw & \qw & \qw & \qw & \qw & \rstick{\ket{a_{n-2}}}\\
       \lstick{\vdots} & & & & & & & & & & \rstick{\vdots} \\
       \lstick{} & & & & & & & & & & \\
       \lstick{\ket{a_0}} & \qw & \qw & \qw & \qw & \qw & \qw & \ctrl{1} & \qw & \qw & \rstick{\ket{a_0}} \\
       \lstick{\ket{b_{n-1}}} & \multigate{5}{\texttt{QFT}} & \gate{\texttt{SR 0}} & \multigate{3}{\texttt{SR 1}} & \qw & \qw & \qw & \multigate{5}{\texttt{SR (n-1)}} & \multigate{5}{\texttt{QFT}^{-1}} & \qw & \rstick{\ket{a_{n-1} + b_{n-1}}} \\
       \lstick{} & & & & & \dots & & & & \\
       \lstick{\ket{b_{n-2}}} & \ghost{\texttt{QFT}} & \qw  &  \ghost{\texttt{SR 1}} & \qw & \qw & \qw & \ghost{\texttt{SR (n-1)}} & \ghost{\texttt{QFT}^{-1}} & \qw & \rstick{\ket{a_{n-2} + b_{n-2}}} \\
       \lstick{\vdots} & & & & & & & & & & \rstick{\vdots} \\
       \lstick{} & & & & & & & & & & \\
       \lstick{\ket{b_0}} & \ghost{\texttt{QFT}} & \qw & \qw & \qw & \qw & \qw & \ghost{\texttt{SR (n-1)}} & \ghost{\texttt{QFT}^{-1}}  & \qw & \rstick{\ket{a_0 + b_0}} 
       }
  \end{array}
   \]
     \caption{Quantum QFT-Based Adder Circuit}\label{fig:qadder}
   \end{minipage} 
   \hfill\hfill
 \begin{minipage}[b]{0.38\textwidth}
 \centering
 \begin{tabular}{c@{$\;\;=\;\;$}c}
   \begin{minipage}{0.3\textwidth}
   { \tiny
\Qcircuit @C=0.5em @R=0.5em {
     \lstick{} & \qw     & \multigate{4}{\texttt{SR m}} & \qw & \qw \\
     \lstick{} & \qw     & \ghost{\texttt{SR m}}           & \qw & \qw \\
     \lstick{} & \vdots & & \vdots & \\
     \lstick{} & & & & \\
     \lstick{} & \qw     & \ghost{\texttt{SR m}}           & \qw  & \qw
     }
     }
   \end{minipage} & 
   \begin{minipage}{0.3\textwidth}
   {\tiny
   \Qcircuit @C=0.5em @R=0.5em {
     \lstick{} & \qw     & \gate{\texttt{RZ (m+1)}} & \qw & \qw \\
     \lstick{} & \qw     & \gate{\texttt{RZ m}}          & \qw & \qw \\
     \lstick{} & & \vdots & & \\
     \lstick{} & & & & \\
     \lstick{} & & & & \\
     \lstick{} & \qw     & \gate{\texttt{RZ 1}}           & \qw  & \qw
     }
     }
   \end{minipage} 
 \end{tabular}
 \caption{\texttt{SR} unfolds to \texttt{RZ} gates.}
 \label{fig:sr-meaning}
 \end{minipage}
   \end{figure}
  
 A QFT-based adder (\Cref{fig:qadder}) performs addition differently than a ripple-carry adder.
 It usually comes with two qubit arrays $y$ and $u$, tries to sum the $y$ bits into the $u$ array, by first transforming $u$'s qubits to QFT-basis and performing addition in the basis, i.e., instead of performing bit arithmetic in a ripple-carry adder, it records addition results via phase rotations.
 The final inversed QFT operarion $\cn{QFT}^{-1}$ transforms the addition result in the qubit phase back to basis vectors.
 We show the distributed version of a QFT-adder below, which has a different way of distribution than the ripple-carry adder above.

\begin{example}[Distributed QFT Adder]\label{def:example7}\rm 
We define the adder as the membrane definition below. Membrane $l$ holds qubit array $x$ and membrane $r$ takes care of qubit array $y$, and they share two $n$-qubit quantum channels $c$ and $c'$. $\cn{C-SR}(j)$ is the controlled \cn{SR} operation, where $\ssassign{x[j]\sqcupplus y[0,n)}{}{\cn{C-SR}(j)}$ means controlling over $x[j]$ on applying \cn{SR} to the $y\srange{0}{n}$ range.

 {\small
 \begin{center}
 $
 \begin{array}{l}
 \textcolor{blue}{\text{Recursive Combinator: }}
 \\
 \begin{array}{c}
 Rec(j,n,f) = \sifb{j = n}{0}{\sact{f(j)}{Rec(j\splus 1)}}
 \end{array}
 \\
 \textcolor{blue}{\text{Process Definitions: }}\\
 \begin{array}{ll}
 Se(j) = \sact{\sact{\downd{c(1)}}{\csenda{c(1)}{x[j]}}}{\sact{\downd{c'(1)}}{\creva{c'(1)}{y}}} & SeR(n) = Rec(0,n,Se)\\
 Re(j) = \sact{\sact{\downd{c(1)}}{\creva{c(1)}{w}}}{\sact{\ssassign{w\sqcupplus y[0,n)}{}{\cn{C-SR}(j)}}{\sact{\downd{c'(1)}}{\csenda{c'(1)}{w}}}}
 &
 ReR(n) = Rec(0,n,Re)
 \end{array}
 \\
 \textcolor{blue}{\text{Membrane Definition: }}
 \\
 \begin{array}{c}
 \qquad\qquad
{\scell{SeR(n)}_l},\,
 {\scell{\sact{\ssassign{y[0,n)}{}{\cn{QFT}}}{\sact{ReR(n)}{\sact{\ssassign{y[0,n)}{}{\cn{QFT}^{-1}}}{0}}}}_r}
 \end{array}
 \end{array}
$
 \end{center}
 }
 \end{example}

 In the above example,Membrane $r$ transforms qubit array $y$ to be in QFT-basis. Each loop step in $SeR$ and $ReR$, we create two  quantum channels ($c(1)$ and $c'(1)$). 
 Membrane $l$ sends a qubit in the $x$ array at a time to membrane $r$ via the channel $c(1)$.
 In the $j$-th iteration, membrane $r$ receives the information in the qubit $x[j]$, stored as $\glocus{x[j]}{r}$,
 and applies a $\cn{C-SR}$ operation that controls over the qubit $\glocus{x[j]}{r}$ on applying $\cn{SR}$ operation on the $y$ qubit array.
 Assume that the qubit state in $x[j]$ is $\ket{d_j}$ ($d_j = 0$ or $d_j = 1$), the controlled $\cn{SR}$ operation adds $2^j * d_j$ to array $y$'s phase by performing a series of \cn{RZ} rotations.
 Then, we teleport $x[j]$ back to membrane $l$ via another single qubit quantum channel in $c'(1)$.
 After the loop, we apply an inversed \cn{QFT} gate to transform the addition result in $y$'s phase back to its basis vectors.

 In each integration, after membrane $l$ teleports qubit $x[j]$ to membrane $r$, as well as membrane $r$ teleports qubit $c(1)$ to $c'(1)$ in membrane $l$,
 the quantum channel states $\textcolor{spec}{\glocus{c[0]}{l}\sqcupplus \glocus{c[0]}{r}}$ and $\textcolor{spec}{\glocus{c'[0]}{l}\sqcupplus \glocus{c'[0]}{r}}$ are destroyed, so the qubit numbers in membranes $l$ and $r$ are always less than $n$ and $n \splus 1$, respectively.

\subsection{Distributed Hidden Subgroup}\label{sec:hidden}

Quantum programs are probabilistic, and some programs might utilize the nature.
One such example is the repeat-until-success scheme, where the success of a quantum program component execution depends on the success of the observation of a measurement result.
In the hidden subgroup algorithm for an additive group $\mathbb{Z}_m$, it is required to prepare a quantum superposition state $\frac{1}{\sqrt{m}}\sum_{j=0}^{m}\ket{j}$ (note: $m \le 2^n$ might not be $2^n$).
The equivalence checking of the distributed and sequential versions of this kind of program might introduce additional difficulties; such a check can be handled by the \disq observable simulation.
We first examine the distributed hidden subgroup algorithm below ,and then examine the equivalence.

\begin{example}[The State Preparation of Hidden Subgroup]\label{def:example4}\rm
We implement the distributed hidden subgroup algorithm as program ${\scell{R}_l}\cn{,}{\scell{\seq{\creva{c(n)}{w}}{T'}}_r}$,
where the superposition preparation process as process $R$ below. $x[0,n) <m \,\cn{@}\, y[0]$ compares every basis-vector in $x[0,n)$ with $m$ and stores the result in $y[0]$. $\seq{\csenda{c(n)}{x[0,n)}}{\zero}$ teleports qubits from membrane $l$ to $r$, and $T'$ carries the rest of the computation of the hidden subgroup algorithm in membrane $r$. 
We assume an $n$ qubit width quantum channel $c$ as: $\varphi=\textcolor{spec}{\bigcupdot_{i=0}^{n\sminus 1}\{\glocus{c[i]}{l}\sqcupplus \glocus{c[i]}{r} : \sum_{d=0}^1{\frac{1}{\sqrt{2}}\ket{{d}}\ket{d}} \}}$.

{\small
$
\begin{array}{l@{\;}c@{\;}l}
R&=&\seq{\downd{x(n)}}{\seq{\downd{y(1)}}{\sact{\ssassign{x[0,n)}{}{\cn{H}} }{R'}}}
\\
R' &=&\sact{\ssassign{x[0,n)\sqcupplus y[0]}{}{x[0,n) <m \,\cn{@}\, y[0]}}{R''}
\\
R'' &=& \sact{\smea{d}{y[0]}}{\sif{d=0}{R}{\seq{\csenda{c(n)}{x[0,n)}}{\zero}}}
\end{array}
$
}
\vspace{0.5em}

We show the execution transitions below.

{\tiny
\begin{center}
$
\begin{array}{c@{\;\;}c@{\;\;}l}
&&
(\textcolor{spec}{\varphi},{\scell{R}_l}\cn{,}{\scell{\seq{\creva{c(n)}{w}}{T'}}_r})
\\[0.3em]
\textcolor{blue}{(1)}
&
\xrightarrow{l.1}
&
(\textcolor{spec}{\varphi\cupdot\{\glocus{x[0,n)}{l}:\ket{\overline{0}}\}},{\scell{\seq{\downd{y(1)}}{\sact{\ssassign{x[0,n)}{}{\cn{H}} }{R'}}}_l}\cn{,}{\scell{\seq{\creva{c(n)}{w}}{T'}}_r})
\\[0.3em]
\textcolor{blue}{(2)}
&
\xrightarrow{l.1}
&
(\textcolor{spec}{\varphi\cupdot\{\glocus{x[0,n)}{l}:\ket{\overline{0}}\}\cupdot\{\glocus{y[0]}{l}:\ket{0}\}},{\scell{{\sact{\ssassign{x[0,n)}{}{\cn{H}} }{R'}}}_l}\cn{,}{\scell{\seq{\creva{c(n)}{w}}{T'}}_r})
\\[0.3em]
\textcolor{blue}{(3)}
&
\xrightarrow{l.1} 
&
(\textcolor{spec}{\varphi \cupdot \{\glocus{x[0,n)}{l}:\frac{1}{\sqrt{2^n}} \sum_{j=0}^{2^n-1} \ket{j}\} \cupdot \{ \glocus{y[0]}{l}:\ket{0}\}},{\scell{{R'}}_l}\cn{,}{\scell{\seq{\creva{c(n)}{w}}{T'}}_r})
\\[0.3em]
\textcolor{blue}{(4)}
&
\equiv
&
(\textcolor{spec}{\varphi \cupdot \{\glocus{x[0,n)}{l}\sqcupplus \glocus{y[0]}{l}:\frac{1}{\sqrt{2^n}} \sum_{j=0}^{2^n-1} \ket{j}\ket{0}\}},{\scell{{R'}}_l}\cn{,}{\scell{\seq{\creva{c(n)}{w}}{T'}}_r})
\\[0.3em]
\textcolor{blue}{(5)}
&
\xrightarrow{l.1} 
&
(\textcolor{spec}{\varphi \cupdot \{\glocus{x[0,n)}{l}\sqcupplus \glocus{y[0]}{l}:\frac{1}{\sqrt{2^n-m}} \sum_{j=m}^{2^n-1} \ket{j}\ket{0}+\frac{1}{\sqrt{m}} \sum_{j=0}^{2^n} \ket{j}\ket{1}\}},{\scell{{R''}}_l}\cn{,}{\scell{\seq{\creva{c(n)}{w}}{T'}}_r})
\\[0.3em]
\textcolor{blue}{(6)}
&
\xrightarrow{l.\frac{m}{2^n}} 
&
(
\textcolor{spec}{\varphi\cupdot\{\glocus{x[0,n)}{l}:\frac{1}{\sqrt{m}}\sum_{j=0}^{m}\ket{j}\}}
,{\scell{\sif{1=0}{R}{\seq{\csenda{c(n)}{x[0,n)}}{\zero}}}_l}\cn{,}{\scell{\seq{\creva{c(n)}{w}}{T'}}_r})
\\[0.3em]
\textcolor{blue}{(7)}
&
\xrightarrow{l.1} 
&
(\textcolor{spec}{\varphi\cupdot\{\glocus{x[0,n)}{l}:\frac{1}{\sqrt{m}}\sum_{j=0}^{m}\ket{j}\}},{\scell{\seq{\csenda{c(n)}{x[0,n)}}{\zero}}_l}\cn{,}{\scell{\seq{\creva{c(n)}{w}}{T'}}_r})
\\[0.3em]
\textcolor{blue}{(8)}
&
\xrightarrow{l.1} 
&
(\textcolor{spec}{\varphi\cupdot\{\glocus{x[0,n)}{l}:\frac{1}{\sqrt{m}}\sum_{j=0}^{m}\ket{j}\}},{\sacell{\seq{\csenda{c(n)}{x[0,n)}}{\zero}}{\emptyset}_l}\cn{,}{\scell{\seq{\creva{c(n)}{w}}{T'}}_r})
\\[0.3em]
\textcolor{blue}{(9)}
&
\xrightarrow{r.1} 
&
(\textcolor{spec}{\varphi\cupdot\{\glocus{x[0,n)}{l}:\frac{1}{\sqrt{m}}\sum_{j=0}^{m}\ket{j}\}},{\sacell{\seq{\csenda{c(n)}{x[0,n)}}{\zero}}{\emptyset}_l}\cn{,}{\sacell{\seq{\creva{c(n)}{w}}{T'}}{\emptyset}_r})
\\[0.3em]
\textcolor{blue}{(10)}
&
\xrightarrow{l.r.1} 
&
(\textcolor{spec}{\varphi\cupdot\{\glocus{x[0,n)}{r}:\frac{1}{\sqrt{m}}\sum_{j=0}^{m}\ket{j}\}},{\scell{{\zero}}_l}\cn{,}{\scell{{T'[x[0,n)/w]}}_r})
\end{array}
$
\end{center} 
}

\end{example}

In the above transitions, steps (1) and (2) create an $n$ qubit array for $x[0,n)$ (as $\cglocus{x[0,n)}{l}$) and a single qubit for $y[0]$ (as $\cglocus{y[0]}{l}$),
and step (3) applies $n$ Hadamard gates to $x[0,n)$, resulting in an $n$-qubit uniformed superposition.
Step (4) rewrite the two qubit groups together into one as a locus $\cglocus{x[0,n) \sqcupplus y[0]}{l}$, while step (5) applies a quantum oracle operation, i.e., a quantum comparison operator.
For every $x[0,n)$'s position basis $\ket{j}$, we check if it is greater than $m$ or not.
This step essentially partitions all the superposition basis-kets into two groups labeled by the $+$ operation in step (5).
The first group contains basis-kets where $j \ge m$ indicated by $y[0]$'s position basis $\ket{0}$, and the second group contains basis-kets where $j < m$ indicated by $y[0]$'s position basis $\ket{q}$. Such a quantum oracle circuit implementation is introduced in \cite{oracleoopsla}.

Step (6) applies a partial measurement operation on $y[0]$ in membrane $l$, with the measurement result $1$.
This results in the basis-kets in $x[0,n)$ collapsing to the second group described above.
Since the total number of different basis-kets in the original uniform superposition is $2^n$, and there are $m$ different choices in the second group.
This means that the measurement probability is $\frac{m}{2^n}$ for measuring out $1$.
This also indicates that we also need to normalize the amplitudes in the $x[0,n)$'s remaining state, and the multiplication factor is $\sqrt{\frac{2^n}{m}}$, the square-root of the inverted number of the probability value $\frac{m}{2^n}$.
This is why the result state amplitude value is $\sqrt{\frac{2^n}{m}} \cdot \frac{m}{2^n} = \frac{1}{\sqrt{m}}$.
The final step above performs a classical conditional.

The above transitions are only one of the possible paths. It is possible that membrane $r$ can perform a nondeterministic step for execution between (1) and (8).
Another possibility is that the measurement in line $(5)$ can measure out $0$, which leads to a repetition of the transitions before (6).
The process demonstrates a repeat-until-success scheme, i.e., we try to generate the correct superposition by conducting measurements, until the correct one, measuring out $1$, appears.

We now show the simulation of the distributed Hidden subgroup program with its sequential version. We first show the sequential program below.

{\small
\begin{center}
$
R_s = \seq{\downd{x(n)}}{\seq{\downd{y(1)}}{\sact{\ssassign{x[0,n)}{}{\cn{H}} }{\sact{\ssassign{x[0,n)\sqcupplus y[0]}{}{x<m \,\cn{@}\, y[0]}}{\sact{\smea{d}{y[0]}}{\sact{\boxed{d}}{\sif{d}{R}{T'}}}}}}}
$
\end{center}
}

We place the synchronization point $\boxed{d}$ after the measurement operation with the measurement result $d$ as the classical value label to compare.
We also need to modify the $R''$ process in the distributed version (\Cref{def:example4}) as $\sact{\smea{d}{y[0]}}{\sact{\boxed{d}}{\sif{d}{R}{\seq{\csenda{c(n)}{x[0,n)}}{\zero}}}}$.
To see how the simulation works, let's skip the transitions from (1) to (6) by omitting the intermediate transitions and focus on the three transition steps after (6), demonstrated as follows.

{\tiny
\begin{center}
$
\begin{array}{cl}
& (\textcolor{spec}{\varphi},{\scell{R}_l}\cn{,}{\scell{\seq{\creva{c(n)}{x}}{T'}}_r})\\[0.2em]
\longrightarrow ... \xrightarrow{l.1.\frac{m}{2^n}}&
(\textcolor{spec}{\varphi\cupdot\{\glocus{x[0,n)}{l}:\frac{1}{\sqrt{m}}\sum_{j=0}^{m}\ket{j}\}},{{\scell{\sact{\boxed{1}}{\sif{1=0}{R}{\seq{\csenda{c(n)}{x[0,n)}}{\zero}}}}_l}\cn{,}{\scell{\seq{\creva{c(n)}{w}}{T'}}_r}})
\\[0.2em]
\xrightarrow{l.1}
&
(\textcolor{spec}{\varphi\cupdot\{\glocus{x[0,n)}{l}:\frac{1}{\sqrt{m}}\sum_{j=0}^{m}\ket{j}\}},{{\sacell{\sact{\boxed{1}}{\sif{1=0}{R}{\seq{\csenda{c(n)}{x[0,n)}}{\zero}}}}{\emptyset}_l}\cn{,}{\scell{\seq{\creva{c(n)}{w}}{T'}}_r}})
\\[0.2em]
\xrightarrow{1.1}
&
(\textcolor{spec}{\varphi\cupdot\{\glocus{x[0,n)}{l}:\frac{1}{\sqrt{m}}\sum_{j=0}^{m}\ket{j}\}},{{\scell{{\sif{1=0}{R}{\seq{\csenda{c(n)}{x[0,n)}}{\zero}}}}{\emptyset}_l}\cn{,}{\scell{\seq{\creva{c(n)}{w}}{T'}}_r}})
\end{array}
$
\end{center} 
}

\begin{wrapfigure}{r}{4.7cm}
\includegraphics[width=0.36\textwidth]{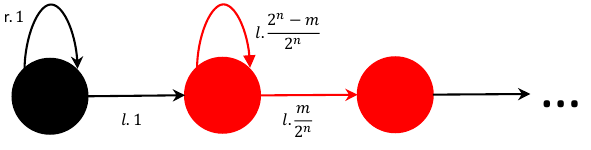}
\caption{Transition Automaton}
\label{fig:hidden-graph}
\end{wrapfigure}

The above transitions can be summarized as the automaton in \Cref{fig:hidden-graph} for highlighting the marked red probabilistic choice components.
Here, only the marked red part happens in the single process $R$ above, and the top-level membrane execution is represented as the root node (marked back on the left)
that has non-deterministic edges choosing $l$ and $r$ for execution.
The $l.1$ edge points to the process-level execution in $R$, representing that we choose to execute the process in $l$.
The self-edge in the marked red node represents the $y[0]$'s measurement resulting in $0$ with a probability $1-\frac{m}{2^n}$, and the measurement of $1$ moves to the next marked red node.
Going through each edge results in a further probability reduction.
For example, every step of measuring out $0$ for $y[0]$ indicates going through the circular edge and results in a $1-\frac{m}{2^n}$ probability reduction along the execution path from the root node to the current state.

Apparently, the sequential program above has a similar transition automaton in \Cref{fig:hidden-graph}.
To simulate the distributed hidden subgroup program with the sequential version, we can classify all the execution paths into two sets: one includes the paths for measuring $y[0]$ to be $0$, and the other contains the paths for its measurement to be $1$.
We can then equate two sets of transitions via the \disq observable simulation.

\begin{theorem}[Distributed Hidden Subgroup Simulation]\label{thm:hidden-sim}\rm 
Let Dis-Hid refer to the distributed Hidden Subgroup program and Hid refer to the sequential one; thus, Dis-Hid $\sqsubseteq$ Hid.
\end{theorem}

We verify \Cref{thm:hidden-sim} in Coq and utilize the same $\cn{not\_sim}$ simulation checking procedure above to automatically validate the theorem.

\subsection{Distributed Quantum Ripple-carry Adders}\label{sec:dis-adder}

\begin{figure}[t]
{
  \includegraphics[width=\textwidth]{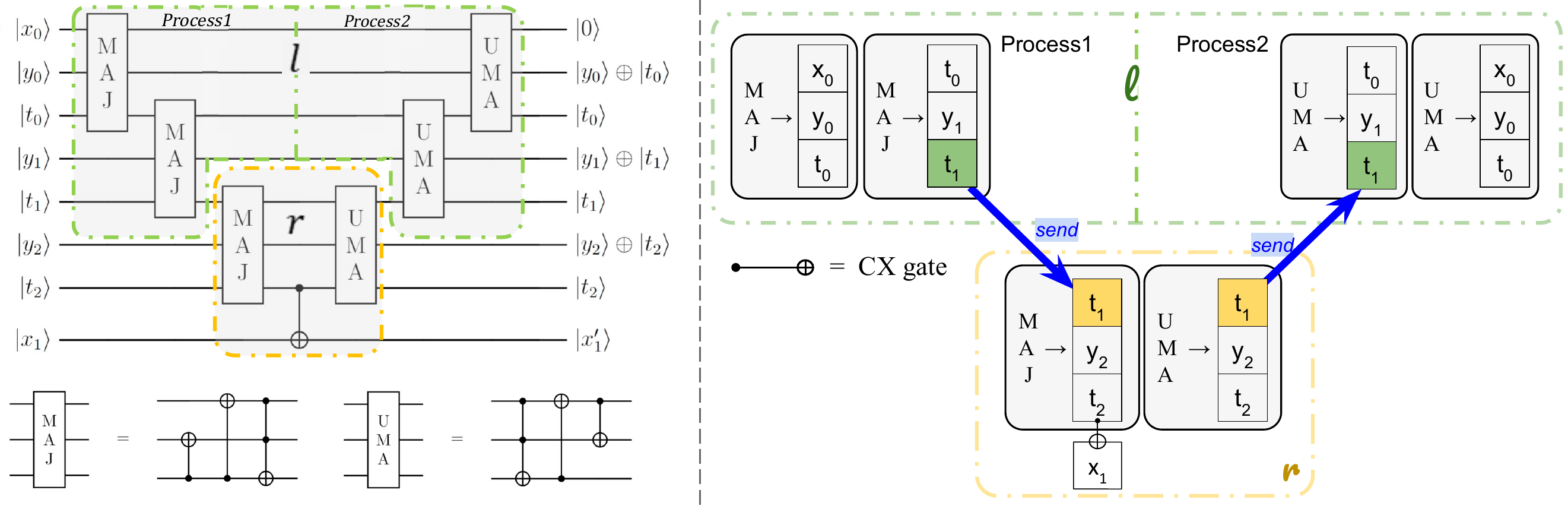}
  \vspace*{-0.7em}
   \caption{Ripple-carry adder. (left): sequential version, (right): distributed version. $x'_1$: overflow bit.}
     \label{fig:adddis}
\label{fig:adds}
}
\end{figure}

Quantum oracle circuits are reversible and used as subroutines in many quantum algorithms; they usually perform the quantum version of some classical computations, e.g., the oracle component in Shor's algorithm is a quantum version of a modulo-multiplication circuit.
They are usually the most resource-consuming component in a quantum circuit \cite{oracleoopsla} and can be implemented as arithmetic operations based on quantum addition circuits.
Distributing the execution of oracle circuits to remote machines can greatly mitigate the entanglement resource needs in a single location.
Here, we show the example of distributing a quantum ripple-carry adder \cite{ripple-carry}. We also describe the distributed QFT-based adders in \Cref{sec:qft-adder}.

\Cref{fig:adddis} (left) shows the sequential circuit of a three-qubit ripple-carry adder, where we add the value of a three-qubit array $t$ to the value stored in the three-qubit array $y$,
with a two-qubit array $x$ storing extra carry qubits, one for the initial carry and the other for an overflow indicator.

{\small
\begin{center}
$
\sact{\ssassign{x[0]\sqcupplus y[0]\sqcupplus t[0]}{}{\cn{MAJ}}}{\sact{\ssassign{x[0]\sqcupplus y[0]\sqcupplus t[0]}{}{\cn{UMA}}}{0}}
$
\end{center}
}

A quantum ripple-carry adder is constructed by a series of \cn{MAJ} operations followed by a series of \cn{UMA} operations, each of which has a diagram on the left side of \Cref{fig:adddis} (left).
To understand the effect of the \cn{MAJ} and \cn{UMA} pairs, we show the application of such a pair to qubits $x[0]$, $y[0]$, and $t[0]$ above.
Here, $x[0]$ is a carry flag for lower significant bits, and $y[0]$ and $t[0]$ are the two bits to add.
The application of the \cn{MAJ} operation adds $t[0]$ to $y[0]$, computes the carry flag for the next significant position, and stores the bit in $t[0]$.
The application of the \cn{UMA} operation reverses the computation in $x[0]$ and $t[0]$ back to their initial bits, but computes the additional result of adding $x[0]$, $y[0]$, and $t[0]$, stored in $y[0]$.
As shown in \Cref{fig:adddis}, we arrange the \cn{MAJ} and \cn{UMA} sequences in the pattern that every \cn{MAJ} and \cn{UMA} pair is placed to connect a carry bit and two bits in the same significant position of arrays $y$ and $t$. The \cn{CX} gate in the middle of the circuit produces the overflow flag stored in $x[1]$. We define these steps in \disq as the following operations.

We distribute the adder to be executed in two membranes, $l$ and $r$, as shown in \Cref{fig:adddis} (right).
Here, we further concurrently execute the two \cn{MAJ}s and \cn{UMA}s, respectively, through two different processes in $l$.
To enable the communication between $l$ and $r$, we utilize our message communication operations. Below, we define the distributed ripple-carry adder, analogous to \Cref{fig:adddis} (right).

\begin{figure}[t]
{\hspace*{2em}
  \includegraphics[width=0.75\textwidth]{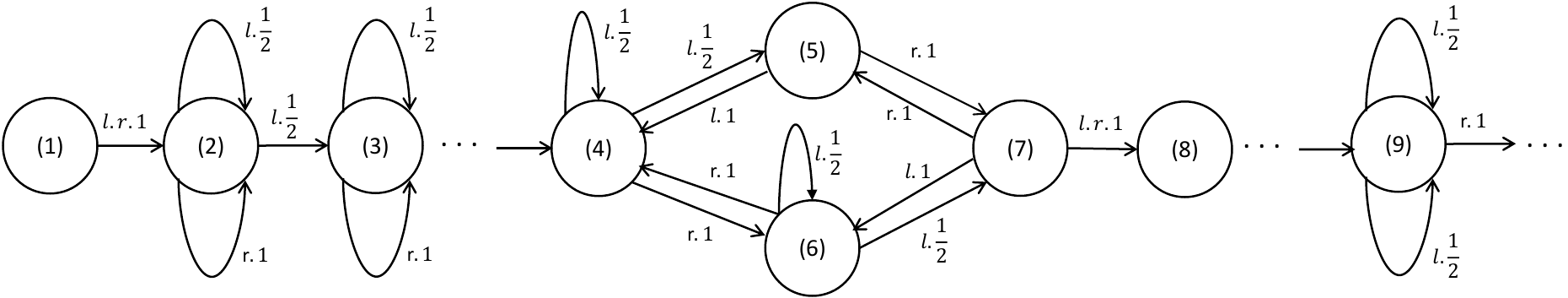}
}
   \caption{The adder automaton.}
   \vspace*{-0.7em}
\label{fig:adder-auto}
\end{figure}

\begin{example}[Distributed Ripple-Carry Adder]\label{def:example6}\rm The following program represents a $3$-qubit distributed ripple-carry addition circuit and has two membranes $l$ and $r$.
Qubits $x[0]$, $y[0,2)$, and $t[0,2)$ belong to membrane $l$, and qubits $x[1]$, $y[2]$, and $t[2]$ belong to membrane $r$.
Qubit arrays $y$ and $t$ are the input qubits storing two $3$-qubit bitstrings as numbers, $y$ stores the final output of adding the two numbers, and $x[0]$ is an ancilla initial carry qubit, $x[1]$ stores the overflow bit.

{\scriptsize
\begin{center}
$\hspace*{-0.9em}
\begin{array}{l}
\scell{\sact{\ssassign{x[0]\sqcupplus y[0]\sqcupplus t[0]}{}{\cn{MAJ}}}{\sact{\ssassign{t[0]\sqcupplus y[1]\sqcupplus t[1]}{}{\cn{MAJ}}}{\sact{\sact{\downd{c(1)}}{\csenda{c(1)}{t[1]}}}{0}}},
\\[0.2em]\qquad\qquad\qquad\qquad\qquad\qquad
{\sact{\sact{\downd{c'(1)}}{\creva{c'(1)}}{w}}{\sact{\ssassign{t[0]\sqcupplus y[1]\sqcupplus w}{}{\cn{UMA}}}{\sact{\ssassign{x[0]\sqcupplus y[0]\sqcupplus t[0]}{}{\cn{UMA}}}{0}}}}}_l,
\\[0.6em]
{\scell{\sact{\sact{\downd{c(1)}}{\creva{c(1)}{w}}}{\sact{\ssassign{w\sqcupplus y[2]\sqcupplus t[2]}{}{\cn{MAJ}}}{\sact{\ssassign{t[2]\sqcupplus x[1]}{}{\cn{CX}}}{\sact{\ssassign{c[0]\sqcupplus y[2]\sqcupplus t[2]}{}{\cn{UMA}}}{\sact{\sact{\downd{c'(1)}}{\csenda{c'(1)}{w}}}{0}}}}}}_r}
\end{array}
$
\end{center}
}
\end{example}

In this program, membranes $l$ and $r$ represent different quantum computers. We assume each permits an entanglement of maximal $6$ qubits, which means that each computer is not enough to execute the three-qubit adder, requiring $8$ qubits for execution, so they need to collaborate in executing the adder.
We utilize the first process in membrane $l$ to compute the two \cn{MAJ} applications to $y$ and $t$, then teleport $t[1]$ to membrane $r$ to compute the addition of the third qubits ($y[2]$ and $t[2]$). The teleportation relies on the quantum channel $\textcolor{spec}{\glocus{c[0]}{l} \sqcupplus \glocus{c[0]}{r}}$ and stores $t[1]$'s information in $\cglocus{t[1]}{r}$.
Membrane $r$ operates $t[1]$ and teleports the result state back to membrane $l$, via the quantum channel $\textcolor{spec}{\glocus{c'[0]}{l} \sqcupplus \glocus{c'[0]}{r}}$.
Thereby, $t[1]$'s information is back to $\cglocus{t[1]}{l}$, where the remaining \cn{UMA} operations are applied.
In the two teleportations, the channels $c(1)$ and $c'(1)$ are consumed, so the total number of qubits used in every given time of a membrane is $< 6$.

To show the equivalence between the sequential ripple-carry adder and its distributed version, we have the following proposition.
Since the \disq simulation requires sequence points of measurements, we assume that the sequential adder and its distributed version are extended with measurement operations at the end to measure all qubits.

\begin{theorem}[Distributed Addition Simulation]\label{thm:adder-sim}\rm 
Let Dis-Adder refer to the distributed ripple-carry adder program in \Cref{fig:adddis} and Adder refer to the sequential ripple-carry adder algorithm in \Cref{fig:adddis} (left); thus, Dis-Adder $\sqsubseteq$ Adder.
\end{theorem}

To understand the simulation in \Cref{thm:adder-sim}, we need to understand the probabilistic transitions in the distributed adder, shown as an automaton in \Cref{fig:adder-auto}.
The step (1) creates a two-qubit quantum channel in membranes $l$ and $r$.
The label $l.r.1$ means that we make a non-deterministic choice in $l$ and $r$ with a probability $1$, referring to only one way of making the channel creation.
The (2) transition step has three possibilities. The transitions in the second process in $l$ (having a label $l.\frac{1}{2}$) and membrane $r$ (having a label $r.1$) represent airlocks on membranes $l$ and $r$, respectively, but the airlocks are message receiving operations that are not available at this point; thus, the next very next steps of the two transitions can only perform releasing the airlocks through \rulelab{S-Rev}. This is why two self-edges point to (2) in \Cref{fig:adder-auto}.
The only transition, pushing step (2) to step (3) in the automaton, is the execution of the first process in membrane $l$ (\Cref{fig:adddis}) to execute an \cn{MAJ} operation.
The label $l.\frac{1}{2}$ means that the transition is one of two possible choices in membrane $l$.
The same situation happens in step (3), as an \cn{MAJ} operation in the first process in $l$ can push the automaton towards the next step.

Steps (4) to (8) in the automaton represent the procedure that passes a classical message from membrane $l$ to $r$.
In step (4), $l$'s second process is still waiting to receive a message, but $l$'s first process and membrane $r$ can perform two airlocks, representing that classical communication can be established between the two. Depending on which of the two airlocks performs first, we can transition to either (5) or (6) for performing one of the airlocks, followed by edges from (5) and (6) to (7), indicating the other airlock transition.
Since airlocks can be released, we have backward edges from (7) to (5) and (6) and edges from (5) and (6) to (4).
The transitions from (7) to (8) commit the message-passing communication between membranes $l$ and $r$.
Transition (9) performs a local action in membrane $r$. At this point, the prefixed actions in the two processes in membrane $l$ do not change program states,
i.e., the first process in $l$ is $0$, possibly performing \rulelab{S-Self}, and the second process is waiting to receive a classical message from membrane $r$.
Therefore, we have two self-edges in (9) labeled with $l$.

The simulation of the sequential and distributed adders' program transitions equates to two sets of program states reaching the same states before measurements.
Other than the Coq proof, we perform an automated validation in our Java simulation checker, via the $\cn{not\_sim}$ algorithm in \Cref{sec:disq-equiv}.
In each node in the transition automata, e.g., \Cref{fig:adder-auto}, we collect the set of nodes for the next possible moves, with the validation of equating the label values on the two sides of the simulation.
Quantum data are represented as symbolic values in our checker, and we validate the equivalence of two quantum data by performing property-based testing with many randomly generated assignments for the symbolic values to check the validity of the logical equivalences of quantum data predicate representations.

\section{Quantum Channels and Quantum Teleportation}\label{sec:msgpassing}

We then show the utility of implementing the quantum message passing operations via quantum teleportation.

We demonstrate an example utility of the simulation relation defined in \Cref{def:osim}, for equating the effect of quantum teleportation and quantum communication via a quantum channel.
As we mentioned in \Cref{sec:intro}, local qubits in a single-location processor, modeled by a membrane, cannot be directly referenced by another processor, and two processors require a quantum channel to communicate a qubit of information.
In the quantum teleportation example, we assume a quantum channel $\textcolor{spec}{\glocus{c[0]}{l}\sqcupplus \glocus{c[0]}{r}}$ is given.
Once a quantum channel is established, we can utilize quantum teleportation to transmit the qubit information from one to the other. To illustrate the teleportation strategy, we will discuss the processes $T$ and $R$ below.

\begin{definition}[Quantum Teleporation Processes]\label{def:example0}\rm
We show the two processes of quantum teleportation, with example transitions in \Cref{sec:controled-ghz}.
The $T$ and $R$ processes below might be placed in two different membranes $l$ and $r$, as it teleports the quantum information in $\textcolor{spec}{\glocus{x[0]}{l}}$ to $\textcolor{spec}{\glocus{c[0]}{r}}$ via the quantum channel $\textcolor{spec}{\glocus{c[0]}{l}\sqcupplus \glocus{c[0]}{r}}$ having the state $\textcolor{spec}{\sum_{d=0}^1 \frac{1}{\sqrt{2}}\ket{d}\ket{d}}$. We insert the synchronization points at the end of each process ($\zero$).

{\small
\begin{center}
$
\begin{array}{lcl}
T & = & \sact{\ssassign{x[0]\sqcupplus c[0]}{}{\cn{CX}}}
   {\sact{\ssassign{x[0]}{}{\cn{H}}}{{\sact{\smea{b_1}{c[0]}}
       {\sact{\smea{b_2}{x[0]}}
         {\sact{\csenda{a}{b_1}}{\sact{\csenda{a}{b_2}}{{\zero}}}}}}}}

\\
R &=& \sact{\creva{a}{b_1}}{
     \sact{\creva{a}{b_2}}{
       \sact{\sifc{b_1}{\ssassign{c[0]}{}{\cn{X}}}}
         {\sact{\sifc{b_2}{\ssassign{c[0]}{}{\cn{Z}}}}{{\zero}}}}}
\end{array}
$
\end{center} 
}

\end{definition}

When executing the two processes in two membranes $l$ and $r$, denoted as ${\scell{T}_l},{\scell{R}_r}$.
In $l$, the applications of \cn{CX} and \cn{H} gates encode the qubit $x[0]$ with the channel $c[0]$, to entangle them.
The two measurements ($\mathpzc{M}$) divides the information in $x[0]$ into two parts: $b_1$ and $b_2$. This information is transferred via classical channels carrying the classical bits $b_1$ and $b_2$. 
On receiving the two bits from membrane $l$, the membrane $r$ restores the quantum information in $x[0]$ by conditionally (depending on $b_1$ and $b_2$) applying \cn{Z} and \cn{X} to $\textcolor{spec}{\glocus{c[0]}{r}}$. After the process, $\textcolor{spec}{\glocus{c[0]}{r}}$ has all the information in $x[0]$.

We demonstrate the equivalence between quantum teleportation and our quantum message passing, via the execution of a quantum teleportation program, ${\scell{T}_l},{\scell{\sdot{R}{ \zero}}_r}$, and a message passing program,  ${\scell{\seq{\csenda{c(1)}{x[0]}}{{\zero}}}_l},$ ${\scell{\seq{\creva{c(1)}{c}}{{\zero}}, \zero}_r}$, performing quantum channel communication.
To demonstrate the probabilistic nature of local membrane parallelism and show that the equivalence can be established under parallel process interleaving, we add a $\zero$ process in membrane $r$.

\begin{wrapfigure}{r}{4.8cm}
\vspace{-2em}
  \begin{center}
    \includegraphics[width=0.36\textwidth]{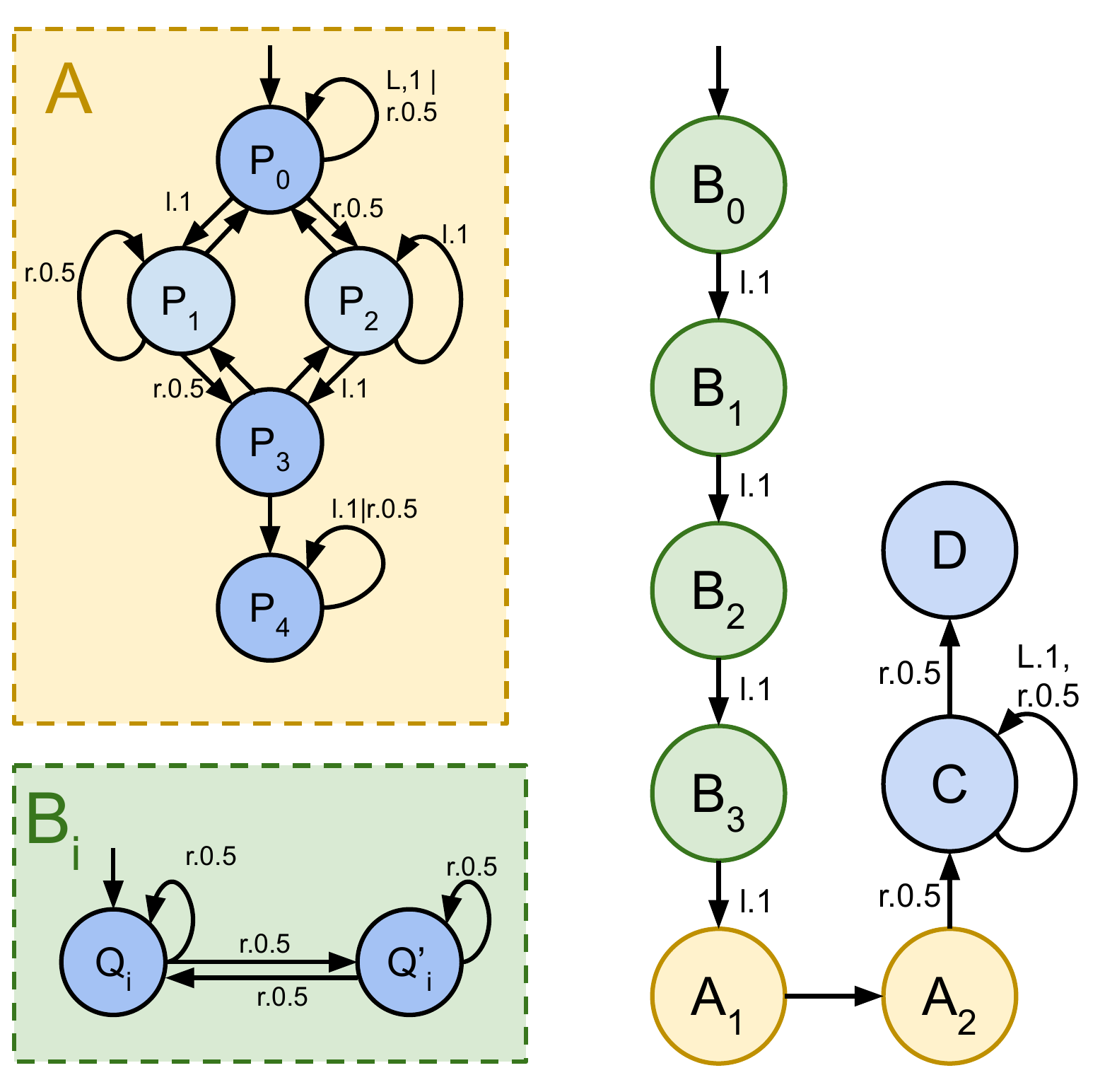}
  \end{center}
  \vspace{-0.5em}
  \caption{Bisimulation of final states of teleportation ($\cn{P}_5$ on the right) and message passing ($\cn{P}_4$ in A)}
  \label{fig:tele-bisim}
\end{wrapfigure}

To demonstrate the equivalence, we can automatonize the two programs in \Cref{fig:tele-bisim}.
The automaton of the message passing program can be represented by \cn{A}, while the quantum teleportation program can be represented by the right-hand automaton, which contains sub-nodes represented by \cn{A} and \cn{B}. The \cn{B} automaton represents the execution of a quantum operation in the process $T$ above, containing two nodes $\cn{Q}_i$ and $\cn{Q}'_i$, representing the possibility of selecting the $\zero$ process in membrane $r$ to execute.
Automaton \cn{A} essentially represents a transition diagram for membrane $r$ to communicate a classical or quantum message from membrane $l$, which is why it can also be used to represent the message passing program.
The equivalence check is conducted by the final quantum states, $\cn{P}_4$ for message passing and $D$ for teleportation, after executing the two programs.

\end{document}